\documentclass[12pt,preprint]{aastex}

\newcommand{\amin}{\arcmin}
\newcommand{\asec}{\arcsec}
\newcommand{\lapprox}{\lesssim}
\newcommand{\gapprox}{\gtrsim}

\shorttitle{X-ray Observations of NGC~1068}
\shortauthors{Colbert et al.}

\begin{document}

\title{The X-ray Reflectors in the Nucleus of the Seyfert Galaxy NGC~1068}

\author{Edward J. M. Colbert\altaffilmark{1,2,3}}
\affil{Johns Hopkins University, Department of Physics and Astronomy, 
Homewood Campus, 3400 North Charles Street, Baltimore, MD~~21218}

\author{Kimberly A. Weaver\altaffilmark{4}}
\affil{Laboratory for High Energy Astrophysics, Code 662, NASA/GSFC,
  Greenbelt, MD~~20771}

\author{Julian H. Krolik}
\affil{Johns Hopkins University, Department of Physics and Astronomy, 
Homewood Campus, 3400 North Charles Street, Baltimore, MD~~21218}

\author{John S. Mulchaey}
\affil{Observatories of the Carnegie Institution of Washington, 813 Santa 
  Barbara Street, Pasadena, CA~~91101}
 
\and

\author{Richard F. Mushotzky}
\affil{Laboratory for High Energy Astrophysics, Code 662, NASA/GSFC,
  Greenbelt, MD~~20771}

\altaffiltext{1}{Laboratory for High Energy Astrophysics, Code 662, NASA/GSFC,
  Greenbelt, MD~~20771}
\altaffiltext{2}{Department of Astronomy, University of Maryland, 
  College Park, MD~~20723}
\altaffiltext{3}{Space Telescope Science Institute, 3700 San Martin Drive, 
  Baltimore, MD~~21218}
\altaffiltext{4}{Johns Hopkins University, Department of Physics and Astronomy, 
Homewood Campus, 3400 North Charles Street, Baltimore, MD~~21218}

\begin{abstract}
Based on observations of the Seyfert nucleus in NGC~1068 
with ASCA, RXTE and BeppoSAX, we report the discovery of 
a flare (increase in flux by a factor of $\sim$ 1.6) in the 
6.7~keV Fe~K line component between observations obtained four months apart, 
with no significant change in the other (6.21, 6.4, and 6.97~keV) 
Fe~K$\alpha$ line components.
During this time, the continuum flux {\it decreased} by $\sim$20\%.
The RXTE spectrum requires an Fe~K absorption edge near 8.6 keV
(Fe XXIII $-$ XXV).
The spectral data indicate that the 2$-$10 keV continuum emission 
is dominated ($\sim$ 2/3 of the luminosity)
by reflection from a previously unidentified region of warm, ionized
gas located $\lapprox$0.2 pc from the AGN.
The remaining $\sim$ 1/3 of the observed X-ray emission is reflected from
optically thick, neutral gas.
The coronal gas in the inner Narrow-Line Region (NLR) and/or the cold
gas at the inner surface of the obscuring ``torus'' are possible cold
reflectors.
The inferred properties of the warm reflector are: size (diameter) 
$\lapprox$0.2 pc,
gas density $n \gapprox$ 10$^{5.5}$ cm$^{-3}$, ionization
parameter $\xi \approx$ 10$^{3.5}$ erg~cm~s$^{-1}$, and 
covering fraction 
0.003 (L$_0$/10$^{43.5}$ erg~s$^{-1}$)$^{-1} < (\Omega/4\pi) <$
0.024 (L$_0$/10$^{43.5}$ erg~s$^{-1}$)$^{-1},$
where L$_0$ is the intrinsic 2$-$10 keV X-ray luminosity of the AGN.
We suggest that the warm reflector
gas is the source of the
(variable) 6.7 keV Fe line emission, and the 6.97 keV Fe line emission.
The 6.7 keV line flare is assumed to be due to an increase in the emissivity
of the warm reflector gas from a {\it decrease} (by 20$-$30\%) 
in L$_0$.
The properties of the warm reflector are most consistent with an intrinsically
X-ray weak AGN with L$_0 \approx$ 10$^{43.0}$ erg~s$^{-1}$.
The optical and UV emission that scatters from
the warm 
reflector into our line of sight is required to suffer strong 
extinction, which can be reconciled if the line-of-sight skims the outer surface
of the torus.
Thermal bremsstrahlung radio emission from the warm reflector may be 
detectable in VLBA radio maps of the NGC~1068 nucleus.
\end{abstract}


\section{Introduction}

NGC~1068 is one of the original ``spiral nebulae'' that were noted by 
Seyfert (1943) to have strong, high-excitation emission lines in their nuclear 
optical spectra.  Its lack of ``broad'' (FWHM $\gapprox$ 1000 km~s$^{-1}$)
emission lines from a ``broad-line region'' (BLR), argued to be present in 
type~1 Seyfert nuclei, identify it as a type~2 Seyfert galaxy.
However, when broad Balmer emission lines were discovered in its 
{\it polarized} 
optical nuclear spectrum by Antonucci \& Miller (1985), it became clear
that this galaxy 
contained
a ``hidden'' broad-line region, seen only in light
reflected around an intervening obscuring medium.  
This result, combined with results for other Seyfert galaxies, 
gave rise to the so-called unified model of active galactic nuclei (AGNs),
which asserts that the difference between type~1 and type~2 AGNs is due
to the relative orientation of an optically thick torus which surrounds
the central engine and BLR (cf. Antonucci 1993 and references therein).

X-ray observations of AGNs have, for the most part, supported this unified
model.  For example, observations of X-ray bright 
type~2 Seyfert galaxies with the
HEAO-1 satellite (Mushotzky 1982) showed that their X-ray spectra were
well represented by a power-law with photon index $\Gamma \approx$ 1.8
(characteristic of type~1 Seyferts), with additional 
absorption with column densities N$_H \gapprox$ 10$^{22}$
atoms cm$^{-2}$.  The absorption column in type~2 Seyfert galaxies, typically
$\sim$10$^{23}$$-$10$^{24}$ cm$^{-2}$, is assumed to be due to the blocking
torus (e.g., Mulchaey, Mushotzky \& Weaver 1992).

Higher quality data now show that Seyfert X-ray spectra are 
more complex than a simple absorbed
power-law.  In many type~2 objects, a `soft excess' (soft X-ray emission in
excess of that expected from a simple absorbed power-law) is common below
$\sim$2 keV (e.g.  Turner \& Pounds 1989).  
Above $\sim$10 keV, the power-law slope generally flattens (e.g.,
Nandra \& Pounds 1994) due to Compton scattering of nuclear light 
reflected from the accretion disk (ergo, the ``reflection component'').
Soft X-ray absorption lines (e.g., due to O, Ne, Mg, Si, Fe) from an
ionized medium are also present (cf. Reynolds \& Fabian 1995).  And finally,
between 6 and 7 keV, strong Fe~K$\alpha$ fluorescence lines 
are present, and are thought to originate from the accretion disk, or, 
in some cases, from the inner edge of the torus.

The central region of NGC~1068 contains a starburst 
(e.g., Wilson et al. 1992), which produces a large fraction of the X-ray
emission below $\sim$4 keV (e.g., Levenson et al. 2001).  At higher
energies, the AGN dominates the spectrum.
The X-ray spectrum of the AGN is particularly interesting, in that
it has an Fe K$\alpha$ line with
a very large equivalent-width (EW), first discovered with 
GINGA by Koyama et al. (1989), who modelled the spectrum with a 
relatively flat ($\Gamma \sim$ 1.5) power-law with little or no
absorption, plus a single Gaussian line centered at 6.55~keV with
EW of 1.3~keV. Compare this with typical EWs for type~1 Seyferts, which are 
$\lapprox$500 eV (e.g., Nandra et al. 1997).  Such a large EW was 
predicted by the X-ray scattering model of Krolik \& Kallman (1987),
although the measured EW in NGC~1068 is larger than predicted.  
It was suggested that the observed 
(reflected) X-ray continuum flux is much smaller than that illuminating the 
fluorescing material, resulting in a larger EW than for Seyfert~1
galaxies.
Higher-resolution X-ray spectral data obtained with BBXRT showed
three separate narrow Fe~K$\alpha$ components 
with a net EW of 2.7 keV (Marshall et al. 1993).  
Marshall et al. suggested that the multiple Fe~K features arose from two 
components of the nuclear gas, a hot and a warm component.  
A revised model of the X-ray continuum for NGC~1068 was later put forward by 
Smith, Done \& Pounds (1993), who found that although a reflection component 
was not required by the GINGA data, if it were included, it would change the 
inferred intrinsic power-law
slope to $\Gamma \sim$ 1.9, more typical of Seyfert~1 X-ray spectra.
More recent
ASCA data have confirmed the three separate line components, this time with 
a total EW of 3.2 keV
(Ueno et al. 1994). 
Using the same ASCA data,
Iwasawa et al. (1997) proposed the existence of a fourth Fe~K component $-$ 
a weak red wing of
the 6.4 keV fluorescence line, possibly
due to Compton scattering in optically thick,
cold matter (Matt et al. 1996).  Hard ($\gapprox$10 keV)
X-ray emission detected by BeppoSAX (Matt et a. 1997) is
consistent with reflection from a mixture of neutral and
ionized gas, with an intrinsic power-law slope of $\Gamma \sim$ 1.7.  

In this paper, we further investigate the nature of the X-ray 
emission due to the AGN by analyzing the broad-band 
(E $\approx$ 4$-$200 keV) X-ray spectrum
using all of the available data from the ASCA, RXTE and
BeppoSAX missions.  We focus here only on the X-ray spectrum
above 4~keV to avoid the nuclear starburst.
The hard ($\gapprox$ 10 keV) X-ray spectral coverage provided by the RXTE and 
BeppoSAX Phoswich Detector System (PDS) 
data is vital for placing constraints on continuum models.  
After the continuum is well modelled with the best data available,
we can then accurately model the Fe~K$\alpha$ complex.  Such
is the goal of the current paper.    
In section 2, we describe the data used for our analyses
and describe details of the data reduction.  Section 3 contains the
primary observational results from our analyses and section 4 contains
a discussion of the implied physical constraints on the geometry of the
nucleus of NGC~1068 and the origin of the X-ray emission.
Throughout this paper, we assume a distance of 
14.67~Mpc (H$_0 =$ 75 km~s$^{-1}$~Mpc$^{-1}$) to NGC~1068.

\begin{figure}[hbn]
\epsscale{0.8}
\plotone{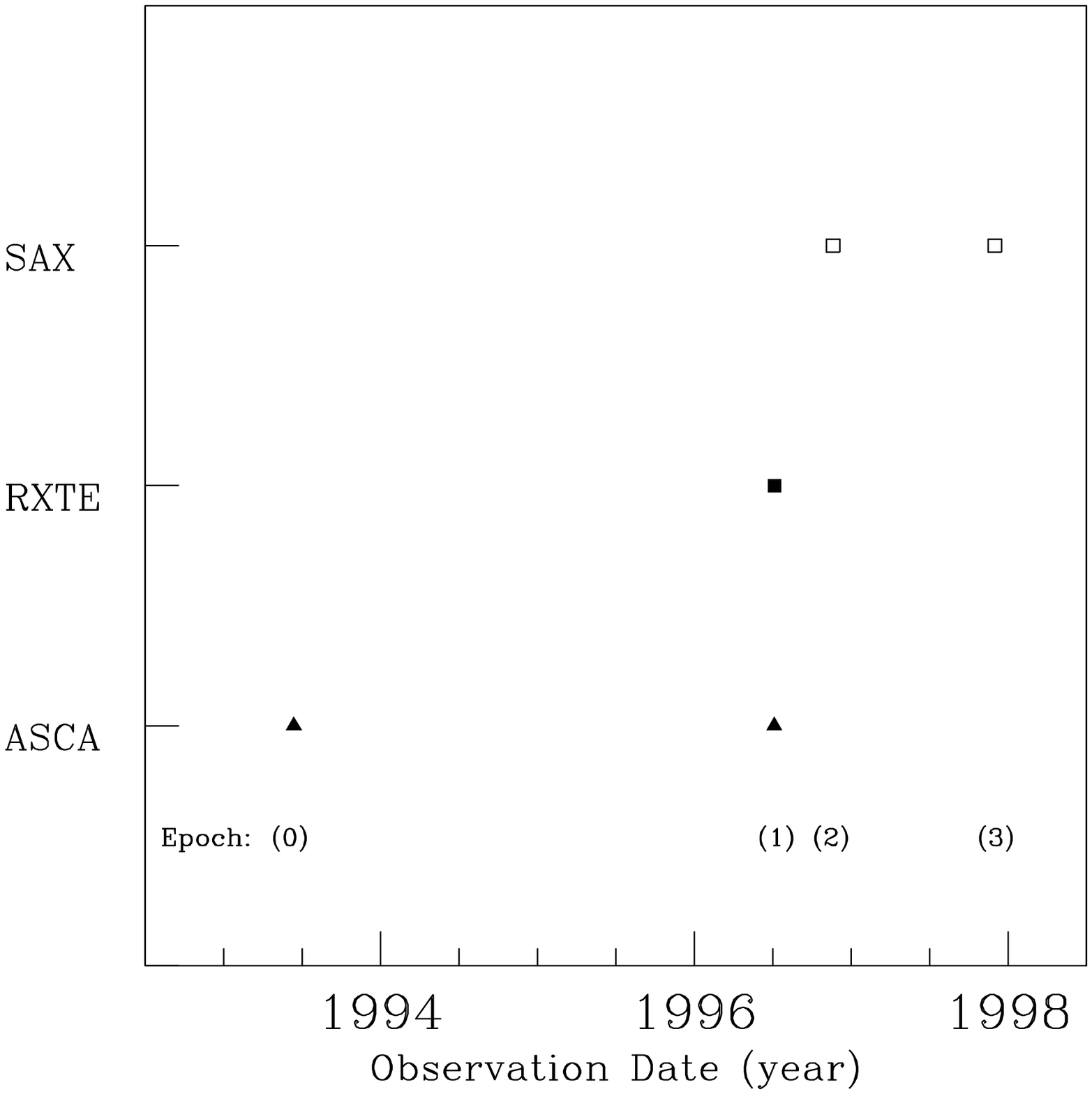}
\caption{
Timeline of relevant X-ray spectral observations of NGC~1068.  
ASCA observations are shown as
filled triangles, RXTE observations are shown as filled rectangles, and 
BeppoSAX observations are shown as open rectangles.  Note that the RXTE
observation is simultaneous with the second ASCA observation (ASCA2).
\label{figallobs}}
\end{figure}

\section{Observations and Data}

We restrict photon energies to be $>$ 4 keV, to avoid any starburst
emission.
A timeline of the relevant spectral observations is shown 
graphically in Figure~\ref{figallobs}, showing the following four epochs:
(epoch~0) ASCA1, (1) ASCA2/RXTE, (2) SAX1 and (3) SAX2.
In this paper, we concentrate mainly on the data from epochs~1 and 2, thus
our odd numbering scheme.
Additional information about the observations are listed in Table~1.

\subsection{ASCA Observations}

Data from the two archival ASCA observations were retrieved from the US 
archives at the NASA/GSFC High Energy Astrophysics Science Archive Research 
Center (HEASARC).  Both the Solid-state Imaging Spectrometers
(SIS0 and SIS1), and the Gas Imaging Spectrometers (GIS2 and GIS3)
were used in our analyses.  The data were screened as follows. 
SIS data data were obtained in bright mode.
Observing times with high background were
removed, based on the recommended criteria in the ASCA data
analysis guide.  Hot and flickering pixels were discarded, as were
data when the satellite passes through the South Atlantic Anomaly (SAA), 
when its elevation above
Earth's limb is $<5^{\circ}$ (night) or $<20^{\circ}$ (day), when the
geomagnetic cutoff rigidity is $<6$~GeV~c$^{-1}$, and when the time is
$<32$~s after a day/night transition, or $<32$~s after an SAA passage.
Background counts were accumulated from source-free regions near the
target and then subtracted.  The ASCA spectra
were grouped to have $\ge30$ photons in each energy bin.

\subsection{RXTE Observations}

The RXTE Proportional Counter Array (PCA) data were taken
simultaneously with the second ASCA observation (ASCA2).
The PCA consists of five Proportional
Counter Units (PCUs), three of which were reliable during the observation.
PCA Standard Mode~2 data were reduced with the
REX (v0.2) script from FTOOLS (v5.0.1), using the default selection
criteria, as follows.  Data were extracted only from layer~1, and only from
PCUs 0, 1 and 2; for Earth elevation angle $>$10$^\circ$, pointing
offset $<$0.02$^\circ$, ``electron contamination'' $<$0.1, and times
$>$30 min after the peak of the last SAA passage.  The background was
estimated using PCABACKEST (v2.1e), using the L7-240 background model,
recommended for faint sources by the NASA/GSFC RXTE Guest Observing Facility.  
The L7-240 model is composed of two components: `L7' and `240.' The `L7'
component estimates the internal detector background derived from a
combination of 7 rates in pairs of adjacent signal anodes, read
directly from the Standard~2 data files.  The `240' component estimates
the effect of particles present in the detectors after an SAA passage,
assuming a decaying exponential with half-life 240~s.  The response
matrix was generated using the FTOOLS task PCARMF (v3.7).  

\subsection{BeppoSAX Observations}

Both sets of BeppoSAX data were retrieved from the archives at the
BeppoSAX Science Data Center (SDC) in Rome.  
Only data from the 
Medium-Energy Concentrator Spectrometers (MECS)  and the 
PDS
instruments were used.
Raw data were processed by 
the SDC's Supervised Standard Science Analysis 
(Rev. 0 for the SAX1 data and Rev. 1.1 for the SAX2 data).
We used BeppoSAX SDC calibration and background data released in November 1998.

The first observation (SAX1) occurred before MECS instrument no.~1
failed (i.e., before May 7, 1997), so events from all three MECS were 
available.  SAX1 MECS spectra were extracted using circular
regions of radius 4\amin\ and then
combined into one spectrum by routine pipeline processing by the BeppoSAX SDC.  
The combined spectrum was then grouped so that it
had a minimum of 20 counts per spectral bin.  Background data
were taken from a 4\amin\ source region from SDC background events files.
The spectral data from the 
PDS instrument were not modified from their original
condition after being processed by the SDC.
The PDS data were grouped 
according to the default grouping file (provided by the SDC).
PDS data processed by the pipeline processing of the BeppoSAX SDC 
are already background-subtracted.

Only two MECS instruments were available during the second SAX
observation (SAX2), since MECS no.~1 had failed by that time.
The SAX2 MECS and PDS data were processed in a similar manner to the SAX1
data, although 
source and background counts were extracted using
circular regions of 3\amin.

\subsection{Notes on Spectral Data}

Our first goal is to
constrain X-ray continuum models of the AGN, so we 
concentrate on the spectral data
with the highest signal-to-noise ratio
and the best broad-band (from $\sim$4 to greater than 10 keV) coverage.
Compton reflection is expected to contribute significantly to the 
continuum shape at energies $\gapprox$ 10 keV, so the ASCA spectra
($\lapprox$10 keV) alone do not constrain this component well.  
However, ASCA data are needed to accurately model the Fe~K$\alpha$ complex.
The RXTE data spectral response (up to $\sim$ 40~keV) is ideally suited 
to model the Compton reflection component, while the
BeppoSAX MECS instrument offers good coverage 
of the Fe~K region and the $\sim$4$-$10 keV continuum.
The BeppoSAX PDS data 
have low resolving power, but are useful in constraining the high-energy X-ray
continuum (up to $\sim$100 keV).  Since the 
SAX1 data have much longer
exposure times than the SAX2 data (Table~\ref{tab1}), 
and were taken nearer to the time of
the epoch~1 data, we use them for modelling the broad-band
X-ray spectrum.  

We therefore initially used the 
simultaneous ASCA2 and RXTE data (epoch~1), and the
SAX1 (epoch~2) data, taken four months later.
XSPEC v11.1.0 is used for all spectral fits.  All errors are quoted at the
90\% uncertainty level for one interesting parameter ($\Delta\chi^2 =$ 2.7).

\section{Spectral Fitting Results}

\subsection{The 4$-$100 keV X-ray Spectrum}

Because the spectrum of NGC~1068 is complex, assumptions about the spectral
structure of the Fe~K region may affect our continuum model.  In this section,
we investigate potential continuum components using four different simple 
continuum models, together with two Fe~K line emission models.  For the line
models, we use four Gaussian components, first an empirical model which 
provides a good fit, and, second, a  
``narrow-line'' model with fixed central energies.  The continuum 
models are: 
(1) an absorbed power-law,
(2) reflection from optically thin ionized gas (power-law plus an Fe~K edge),
(3) reflection from optically thick neutral gas (XSPEC constant-density 
  model {\sc pexrav}; Magdziarz \& Zdziarski 1995), and
(4) reflection from optically thick ionized gas ({\sc pexriv}).
Galactic absorption was included with N$_H$ fixed at 3.53 $\times$
10$^{20}$ cm$^{-2}$ (Dickey \& Lockman 1990).  Relative normalization constants
for each detector were allowed to vary, but were constrained to be within
ranges determined from cross-calibration studies (see section 3.1.3).

\subsubsection{High-energy Continuum \label{hecont}}

In order to
test for variability in the hard continuum between epochs,
we fit the RXTE, SAX1 PDS, and SAX2 PDS
spectra individually with an absorbed power law model.
The 12$-$40 keV flux and the
spectral indices are consistent with no variation between each of the three
observations.  
Therefore, fitting the epoch~2 SAX1 PDS data together with the epoch~1 ASCA 
and RXTE data is justified.

\subsubsection{Empirical Model for the Fe~K complex \label{Feemp}}

For the purposes of finding the best continuum model, 
we first take an empirical approach to modelling the Fe~K$\alpha$ line complex,
using only the epoch~1 data (ASCA2 and RXTE).
Following Iwasawa et al. (1997), we
model the Fe~K line complex with four Gaussian components, but allow
the line energies and line widths to vary.  
After trying the continuum models, we found
that the best fit model was for an optically thick 
ionized reflection continuum model 
($\xi \approx$ 2000 erg~cm~s$^{-1}$)
with Gaussian central energies
(and line widths $\sigma$) at 6.21 (0.01), 6.38 (0.05), 6.59 (0.007), and
6.84 (0.09) keV.  We attach no physical significance 
to these energies and line widths, and note only
that they fit the spectral data well.

We then fixed the central energies, line widths, and line
fluxes to the values from this fit and proceeded to fit the epoch~1 and 2
data together.
The ionized reflection models are again
significantly preferred over the other two 
models ($\chi^2$/dof $=$ 486/419 and 498/419, for the optically
thick and optically ``thin'' 
models, 
compared to 532/420 and 536/420, 
for the absorbed power-law and neutral reflection models, respectively).  
Such a large difference in $\chi^2$ suggests that the dominant component of
the continuum emission is indeed reflection from an ionized plasma.
Both the optically thick and optically
thin ionized reflection models (the best fit models) require 
an edge from ionized Fe at E $\approx$ 8.6 keV (Fe XXIII $-$ XXV).
The best-fit Fe~K edge depth $\tau_{edg}$ is $\sim$0.5 for the 
``optically thin'' model, which is not really in the optically thin or thick
range.  Since there is no
available model for reflection from ionized gas with $\tau \approx$ 0.5 
(see footnote\footnote{
Since $\kappa_{FeK} \approx \kappa_{Th}$ for solar [Fe/H] 
(e.g., Leahy \& Creighton 1993),
$\tau_{edg} \approx \tau_{es}$ 
for $\tau_{es} \lapprox$ 1.
}%
),
in the remainder of the paper, we show results from both our PEXRIV (optically
thick) and power-law-plus-edge (optically thin) models.
\begin{figure}[hbn]
\epsscale{0.8}
\plotone{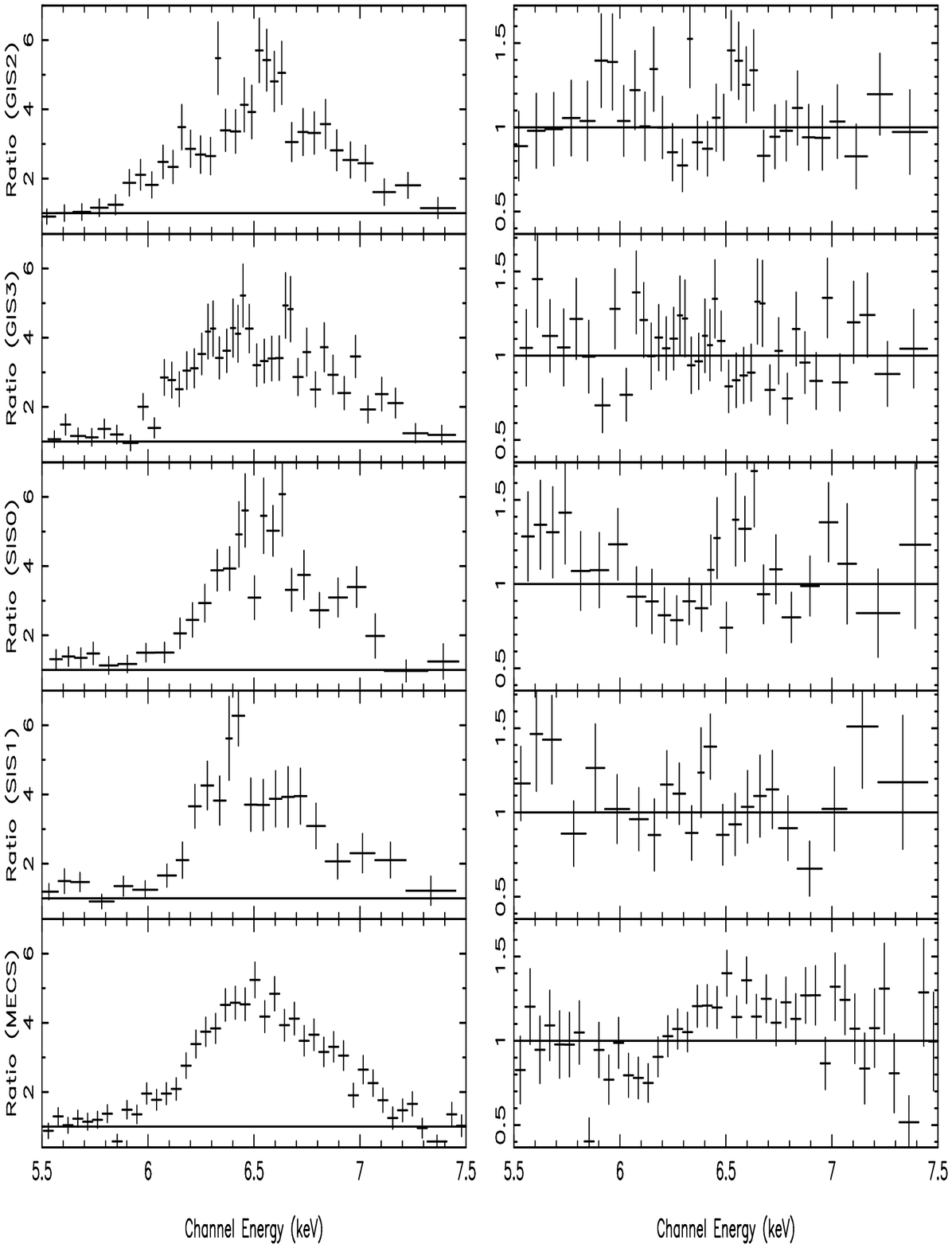}
\caption{
Ratio plot (data to model) of the Fe~K$\alpha$ spectral region for the
absorbed power-law model.  Data from both epoch~1 and epoch~2 
(ASCA2, RXTE and SAX1) were used in this joint fit (see section 3.1.2).
At left, we show the ratio of the epoch~1 ASCA and epoch~2 MECS data to
the power-law model without Fe lines, while at right we show the ratio of the
data to the power-law model with the four-Gaussian model described in 
section 3.1.2.  The top four panels at right show no significant residuals, 
while the epoch~2 MECS ratio plot at bottom right shows
excess emission centered at $\sim$6.7 keV.
\label{figratio}}
\end{figure}

The reduced $\chi^2$ values for the ionized reflection models
($\chi_\nu \approx$ 1.2) further suggest that some features are still
not well modelled.  Indeed, the epoch~2 MECS spectrum shows excess 
emission
centered near 6.7 keV.  In Figure~\ref{figratio}, we show the ratio
of the data to the absorbed power-law model
for energies near the Fe~K complex.
While the
epoch~1 ASCA2 detectors (top four panels) show no 
consistent residual line emission,
residuals are
clearly evident at $\sim$6.7 keV in the epoch~2 MECS spectrum (bottom panel).  
The result is not an artifact of the continuum model, since 
ratio plots using the other continuum models show the same result.
This demonstrates variability in the 
Fe~K$\alpha$ line complex between epochs 1 and 2 (discussed further 
in section 3.2).  For this reason, hereafter we omit 
the SAX1 MECS spectrum in the fit, when fitting for the purpose of 
determining the broad-band continuum shape.

Now, using only the SAX1 PDS data with the ASCA2 and RXTE data 
(and cross-calibration constant ranges described in section 3.1.3),
the fits are much better (see Table 2).  For example, $\chi_\nu^2 =$ 1.0 
for the optically thick ionized reflection model, 
compared to $\chi_\nu^2 =$ 1.3, when the SAX1 MECS is included.
The absorbed power-law model and neutral
reflection models are still not acceptable fits ($\chi^2$/dof =  376/322 and 
401/322, respectively).
The ``optically-thin'' ionized reflection model with an edge at $\sim$8.6 keV
is also a good fit, with
$\chi_\nu \approx$ 1.07.
These results suggest that much of the nuclear X-ray emission
is reflected into our line of sight by highly ionized gas with
$log~(\xi/$~erg~cm~s$^{-1}) \sim$ 3.
We cannot, however, discriminate between an optically thick or thin ionized
reflector.
The Compton reflection component provided by the optically thick model PEXRIV
does give a better fit than the optically-thin reflection model 
($\Delta\chi^2 =$ 31).  This suggests that at least one
optically thick reflector (ionized or neutral; see also Matt et al. 1997)
is present.

\subsubsection{Cross-Calibration of the Instruments}

Relative flux normalization constants between the various ASCA, XTE and SAX 
instruments have been estimated by Yaqoob et al. (2000) and Fiore et al. 
(2001).  The results are that all ASCA instruments and the SAX MECS give
the same flux values within 3\%.   The XTE PCAs give a 20\% higher flux value,
assuming that the are no effects due to their different aperture sizes.
Compared with MECS, the SAX PDS gives a 15\% low flux value, again assuming
no aperture effects.

For the rest of the fits in this paper (the ``narrow-line'' fits), we 
restricted the calibration constants of the instruments to be within our
best-estimated ranges.  This is necessary to insure that any 
variability we find is not due to calibration differences.
We first fixed the calibration of the ASCA GIS2 instrument to 1.0.  
All relative calibration constants of the other instruments are quoted with
respect to GIS2.  We allowed the calibration
constants for all other ASCA instruments and the MECS to vary within a 6\%
uncertainty range (0.94$-$1.06).  The larger aperture of the XTE PCA allows
more flux from the rest of the NGC~1068 galaxy, so we determined the ``best''
PCA normalization constant from the empirical four-Gaussian fit to only
the epoch~1 data, using our best-fit continuum model, the optically-thick 
ionized reflection model.  The
resulting range was 1.34$\pm$6\%, which is consistent with Yaqoob et al.'s
value, given the larger aperture of the PCA.  The ``best'' PDS normalization
constant was determined from fitting the four-Gaussian
empirical models to both the epoch 1 and 2 data.  All continuum models gave
the same result, 0.81, thus we used a range of 0.81$\pm$6\% for the PDS.
This range is also consistent with the estimated value from
the SAX calibration team.

\subsubsection{Narrow-line Fixed-Energy Model for the Fe K complex \label{Fenarrow}}

We next model the Fe~K$\alpha$ line complex with three narrow
($\sigma =$ 0) Gaussian lines and fix the central energies of 
the lines to
those of the expected strong atomic transitions or other observed features:
6.40 keV (`neutral' Fe K$\alpha_1$ and K$\alpha_2$ transitions),
6.70 keV  (He-like Fe XXV), and
6.97 keV  (H-like Fe XXVI).
Adding a line at 6.6 keV (to represent the blend of Fe~XXV and XXIV lines 
expected in a hot plasma, e.g., Beiersdorfer et al. 1993; Feldman, Doschek \& 
Kreplin 1980) does not improve the fit and so we
omit that feature.
However, the
`Compton shoulder' feature at $\approx$ 6.2 keV (Iwasawa et al. 1997) is 
significant, so we include an additional line at 6.21 keV and fix its 
width $\sigma$
to 150 eV.
Line fluxes were allowed to vary.
This narrow-line model is consistent with the Chandra HETG spectra of the
NGC~1068 nucleus published by Ogle et al. (2001), who find that the Fe line
complex is fit very well with unresolved Fe lines at these energies, with
no evidence for a broad Fe line component (P. Ogle, priv. comm.).

Using this narrow-line model,
the ionized reflection continuum
models are again clearly preferred over both the absorbed power-law and neutral
reflection models, as they were for the empirical Fe-line model.
The line fluxes are not significantly different for the four models
(see Table 3), indicating that the line 
parameters are 
not strongly correlated with the choice of continuum model.

\bigskip

In summary, both
the empirical and narrow-line methods of modelling the Fe~K features
imply that the continuum flux from NGC~1068 
includes a 
reflected component from highly ionized gas.
For both optically thick and optically ``thin'' ionized reflection models, the
Fe absorption edge energy is $\sim$8.6 keV, implying that the ionization 
parameter $\xi$ of the warm reflector has a value $log \xi \sim$ 3.
Implications of these results are discussed in section 4.

\subsection{A Variable Fe K line \label{varfek}}

As mentioned in section~\ref{Feemp},
there is excess line emission in the
MECS centered $\sim$ 6.7 keV (Figure~\ref{figratio}),
implying an increase in the epoch-2 6.7 keV line flux, compared to epoch~1.
Here, 
we investigate this variability further by fitting the
narrow-line models discussed above (section~\ref{Fenarrow})
to all of the epoch~1 and epoch~2
data simultaneously.
We perform fits using all four continuum models, for completeness.

\begin{figure}[hbn]
\epsscale{1.0}
\plotone{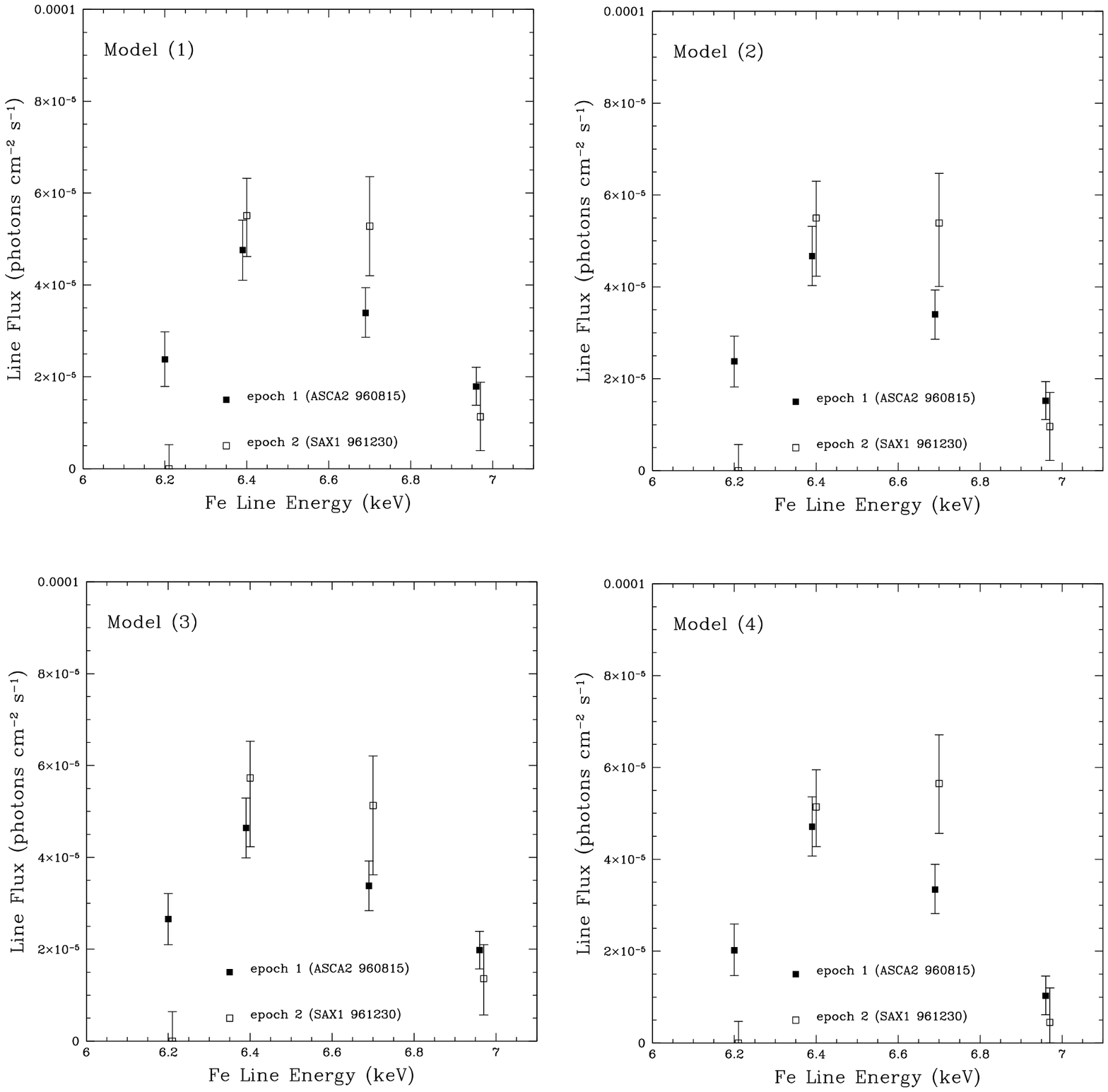}
\vspace*{-1.5in}
\caption{Line fluxes (Table 3) for the four Gaussian Fe~K ``narrow-line''
components 
for epochs~1 and 2, assuming the continuum shape did not vary between epochs,
but allowing the power-law norm (continuum flux) to vary.
The best-fit ionized
reflection models (models 2 and 4) show the main result, which is that
the 6.7 keV line component increased in 4 months.
For completeness, the other two models are shown.  See Table 2 for 
descriptions of continuum models.
Error bars are 90\% confidence ($\Delta\chi^2=2.7$).
\label{figlinvar}}
\end{figure}

First, we froze the the continuum model parameters for both epoch~1 and 2
data to those listed in Table~\ref{tab2}
(the `epoch~1' values), while the four epoch~2
Fe~K line fluxes were allowed to vary.  
Initially, we did not allow the continuum (i.e. power-law normalization) to
vary between epochs.  However, simply allowing the power-law normalization
to vary improved the fit enormously ($\Delta\chi^2 =$ 70$-$80) and 
consistently showed a decrease in the continuum flux by 20\% (uncertainty of
5$-$10\%), independent of the choice of continuum model.  
We discuss the continuum variability further in sections 3.3 and 4.
The resulting line fluxes for the variable-norm case for both epochs are 
listed in Table~\ref{tab3} and are shown
graphically in Figure~\ref{figlinvar}.  Although we do not favor the 
constant-norm case, we also list fluxes for it in Table 3.
Although the ionized reflection continuum models (2 and 4) are 
clearly preferred over
the other two continuum models,
as a precaution, we also performed the same 
test for the absorbed power-law and neutral reflection continuum models.
The 6.7 keV line flux is consistently a factor of $\sim$1.6 higher in epoch~2,
while the 6.4 keV line flux is statistically consistent with no variation.
The larger epoch~2 value for the 6.7 keV line flux
is consistent with measurements from
recent Chandra HETG data (P. Ogle, priv. comm.).

\begin{figure}[hbn]
\epsscale{1.0}
\plotone{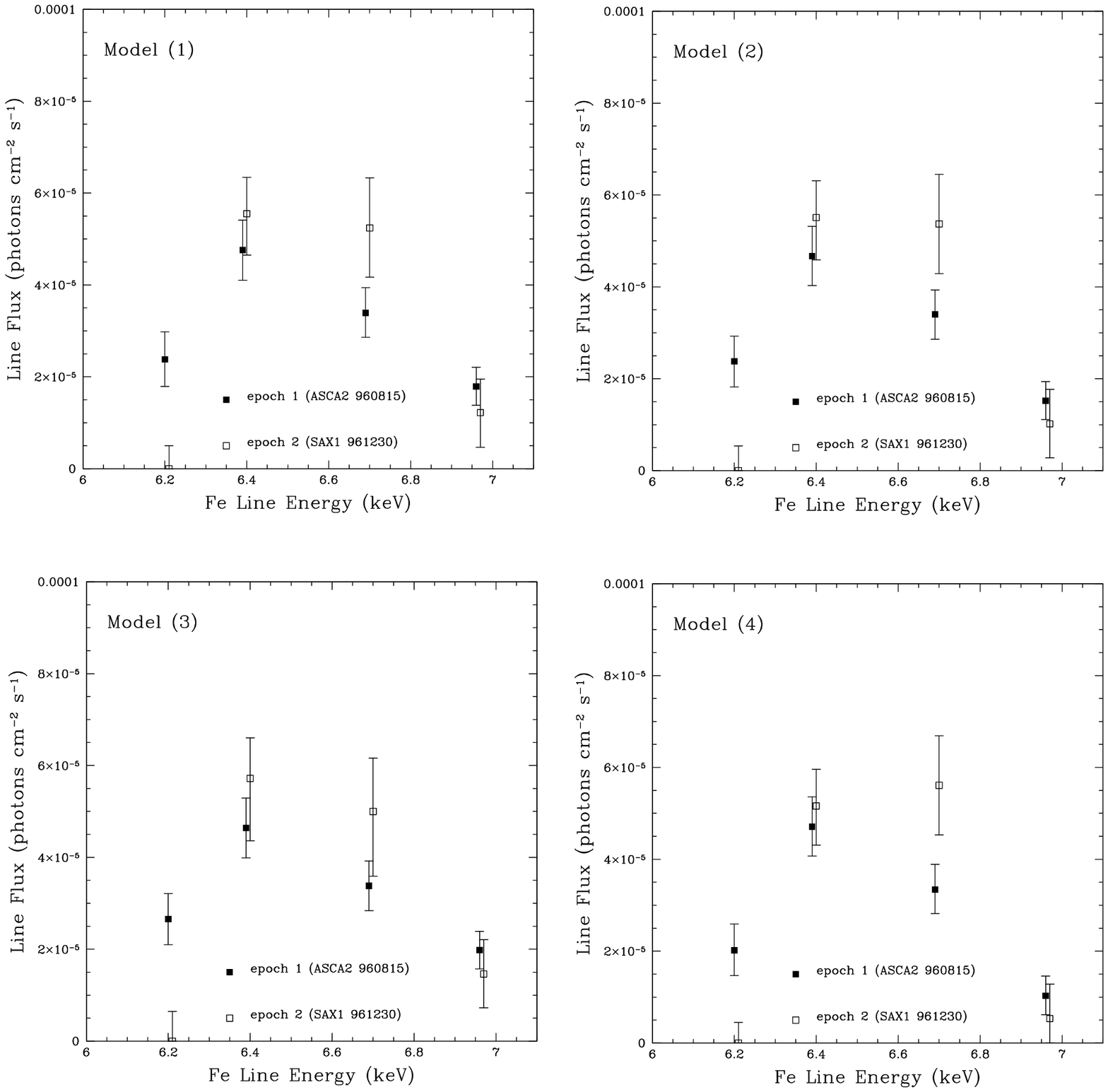}
\vspace*{-1.5in}
\caption{Line fluxes (Table 3) for the four Gaussian Fe~K 
``narrow-line'' components for epochs~1 and 2, when the continuum slope 
$\Gamma$ is allowed to vary between epochs.
Again, the best-fit models (2 and 4) show that the 6.7 keV line increased
between epochs.
\label{figlinvarvf}}
\end{figure}

In order to ensure that the increase in the 6.7 keV line flux was not due to 
a change in the continuum shape, we also allowed
various continuum model parameters to vary between epochs
(see Table 3 and Figure 4).
The main result is that the flux in the 6.7 keV component shows an
increase regardless of the continuum shape.  We should note that
for several cases using the the neutral reflection model, 
while the 6.7 keV line flux does show an increase,
it is not at the 90\% confidence level.
However, we also note that the 
neutral reflection model is a very poor fit to the broad-band spectrum.
The ratio in the line flux (epoch~2/epoch~1)
ranges from 1.53 to 1.75 for the best-fit ionized
reflection models (models 2 and 4).
Thus, we can say with confidence that the structure of the Fe~K line complex
has varied in the four months between the two epochs, and, in particular
the 6.7 keV line flux has increased by a factor of 1.53$-$1.75.
For the case when the continuum spectral shape did not change, but the continuum
flux did, the increase factor 
in the 6.7 keV line flux is 1.6$-$1.7, for the best-fit
ionized reflection models.

We have also attempted to model the Fe~K variability
in other ways.  If the central line energy of the ``6.7~keV'' line is allowed
to vary from 6.70 keV while fitting,
it does not
deviate from 6.70 keV.  If we further allow the line width $\sigma$ to
vary, it increases 
to 0.12 keV.  Even in this case, the 6.7~keV line flux variability is still 
comparable to that for the narrow-line Fe model,
while the other line fluxes are consistent with those in Table~\ref{tab3}.
Thus, even if the line width has increased, the 
line flux has increased as well.

We also fit the Fe~K complex with 6.4 and 6.7 keV lines using a relativistic
disk model (XSPEC model {\sc diskline}) to see if the variability could
be modelled in terms of physical variations in the accretion disk (e.g., 
the inner radius of the accretion disk).  We first tried
replacing only the 6.7~keV Gaussian line with a disk line, and then tried
replacing both
the 6.4 keV line and the 6.7 keV line with a disk line.
However, for both face-on (i $\sim$ 30$^{\circ}$) and edge-on 
(i $\sim$ 70$^{\circ}$) disks, the disk model is a poor fit, regardless of
the central energy. 
We obtain $\chi^2 >$ 500 for a combined fit with epoch~1 and 2 data 
with an ionized reflection continuum model, compared with
$\chi^2 \sim$ 400$-$420 for pure Gaussian line fits in Table~\ref{tab3}.
The fit is poor regardless of what central energy is used for the line and
whether the continuum is allowed to vary between epochs.
Therefore, we conclude that either a relativistic disk line of
this variety is not present,
or, if it is present, the Fe~K complex has too many components
to be able to model it directly with current data.  The recent
Chandra HETG grating spectrum of the NGC~1068 nucleus also shows no evidence
for a broad Fe line component (Ogle et al. 2001), which supports our conclusion
that there is no dominant broad Fe line component present in our data.

\subsection{Long-term Variability of the continuum}

\begin{figure}[hbn]
\epsscale{1.0}
\plotone{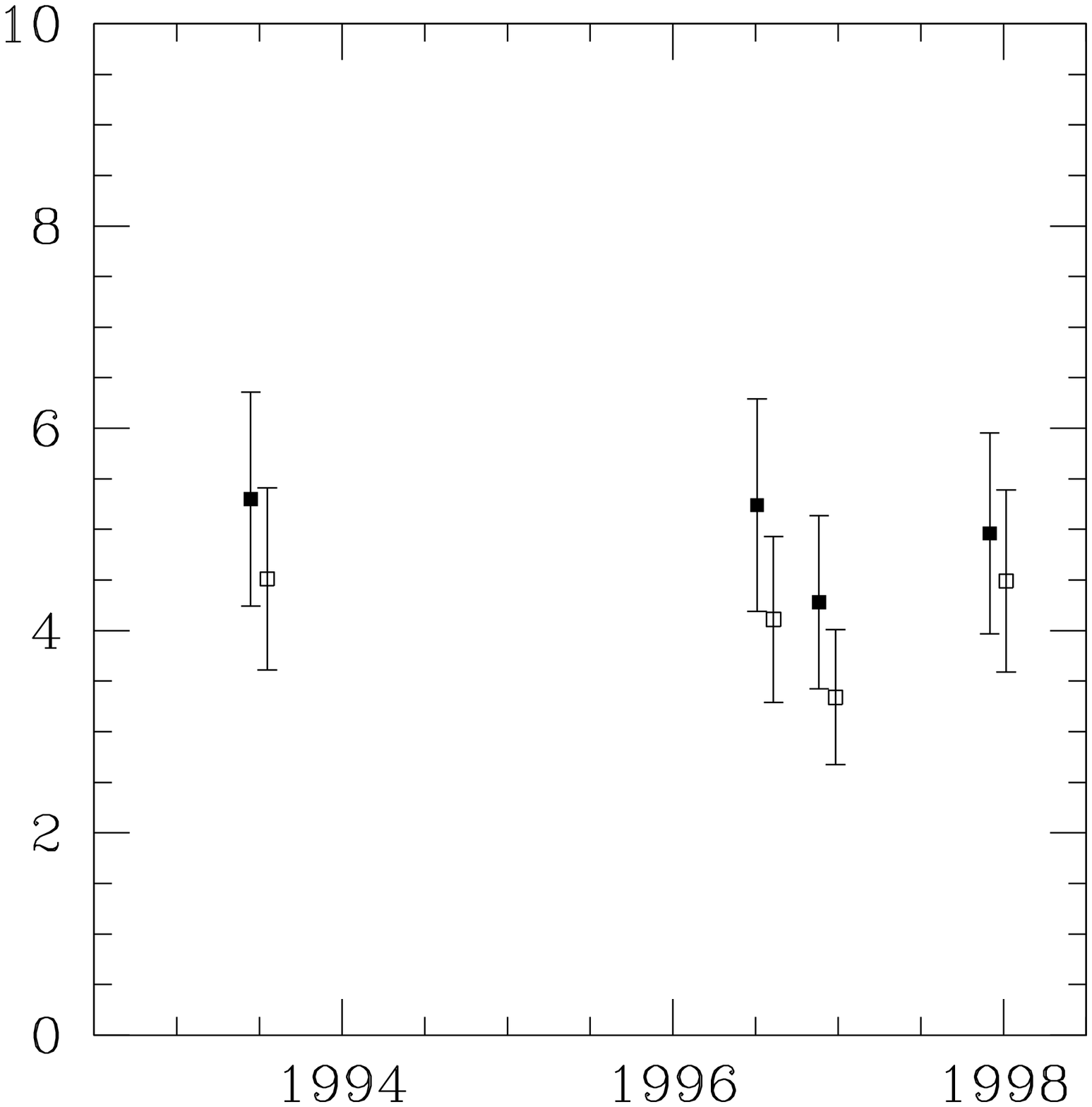}
\caption{
Continuum flux measurements
for four epochs of NGC~1068 X-ray data: (epoch~0) ASCA1, (1) ASCA2, (2) SAX1 
and (3) SAX2.  The ionized reflection continuum model was assumed, but other 
models
give similar results.  The filled rectangles are the 4$-$5 keV flux (no Fe~K
emission) in units 
of 10$^{-13}$ erg~s$^{-1}$~cm$^{-2}$, and the open rectangles are the 4$-$10 
keV flux (including 
the Fe~K emission) in units of 10$^{-12}$ erg~s$^{-1}$~cm$^{-2}$.
The 4$-$10~keV flux points are offset by $+$0.1 yr for clarity.
The plotted uncertainties 
are $\pm$20\%.
\label{figcontvar}}
\end{figure}

We also checked for variability of the continuum flux between all four 
epochs of data.
In Figure~\ref{figcontvar}, we show the 4$-$10 
keV flux (including the Fe~K emission; open rectangles) and the 
4$-$5 keV flux (solid rectangles) from the 
ASCA GIS2 or SAX MECS instruments.
Between the two epochs for which the 6.7~keV component varied (epochs~1 and 2),
the continuum flux decreased  by 18\%, consistent with the 20\% value obtained 
in section 3.2.  The 20\% error bars in the plot correspond with typical
90\% confidence ranges of the power-law norm (``N'' in Table 3) when several
interesting parameters were allowed to vary.
We conclude that the continuum flux
did not vary more than 20\% over the entire timespan (from 1993.5 to 1998),
assuming the continuum shape is allowed to vary.  If the continuum shape
did not vary between epochs 1 and 2, the flux decreased by 
$\approx$20\%, with a subsequent rise in flux between epochs 2 and 3.

It is worth noting that the $\chi^2$ values are consistently lowest, or very
nearly lowest, for the fits in which only the power-law norm and the
Fe~K line fluxes vary between epochs 1 and 2 (see Table 3).  Thus, this 
simple model suggests both an increase in the 6.7 keV line flux and a decrease
in the continuum flux.

\subsection{Summary of Results}

Here we summarize the results from this section.  When the epoch~1
spectral data are fit with Gaussians to model the Fe~K lines, plus the
following continuum models, (1)
an absorbed power-law, (2) an ``optically-thin'' ionized reflection
model (power-law plus edge), (3) an optically-thick neutral reflection
model (PEXRAV), or (4) an optically-thick ionized reflection model
(PEXRIV), the best fit models are (2) and (4).
Both of these models fit the
edge observed in the RXTE data at $\approx$8.6 keV.  
This edge energy implies an
ionization parameter $\xi \sim$ 10$^3$ erg~cm~s$^{-1}$.  Model (4)
gives a better fit than model (2), suggesting that Compton reflection
is needed (from optically-thick ionized gas), or, from an additional
component of optically thick neutral gas.  When the epoch~2 data and
the epoch~1 data (taken 4 months apart) are jointly fit, the continuum
is noted to decrease by $\approx$20\%, while the 6.7~keV Fe~K line
component is observed to increase by a factor of $\approx$1.6.

\section{Discussion}

In the subsections below, we discuss the 
location and physical properties of the X-ray reflection regions in the 
nucleus of 
NGC~1068.  We also discuss the implications for the reflection of light
at other wavelengths.

\subsection{The Intrinsic X-ray Luminosity of the AGN}

The common AGN paradigm predicts that, for type~2 active galaxies, direct
emission from the AGN is partially or fully
blocked by an optically and geometrically thick torus
(cf. Antonucci 1993), so the intrinsic X-ray luminosity L$_0$ cannot be
measured directly.
Even so, we can estimate L$_0$ using
the correlation between the 2$-$10 keV X-ray luminosity
and the [O~III] $\lambda$5007 
optical line luminosity for Seyfert nuclei
(e.g., Mulchaey et al. 1994): log([O~III]/L$_X$) $=$ $-$1.89$\pm$0.25 for
type~1 Seyferts and 
log([O~III]/L$_X$) $=$ $-$1.76$\pm$0.38 for type~2 Seyferts.
We use the correlation for type~1 Seyferts here, since
some of the X-ray emission in type~2 Seyferts is absorbed.
Using the nuclear [O~III] flux from Whittle (1992), we 
estimate the intrinsic 2$-$10 keV X-ray luminosity of the AGN to be 
$L_0 =$ 10$^{43.5}$ erg~s$^{-1}$ (with an uncertainty of $\sim$0.25 in the log).
As a comparison, 
Pier et al. (1994) estimate the intrinsic bolometric luminosity of NGC~1068 
to be 3.8 $\times$ 10$^{44}$ erg~s$^{-1}$
(for a relative reflection fraction f$_{refl} =$0.01).
Assuming L$_X$(2$-$10 keV)/L$_{Bol}$ $\sim$ 0.1 for AGN 
(e.g., Elvis et al. 1994), their estimate of $L_0$ is consistent with that
predicted from the [O~III]/L$_X$ relationship.

\subsection{The 6.7 keV Line Emitting Region}

\subsubsection{Size}

The variability of the 6.7 keV line luminosity in the four
months between epochs 1 and 2 suggests that the maximum size of the 
line-emitting region is approximately the light travel time of 4 light 
months (0.1 pc)\footnote{%
If the variability is due to changes in the density rather than changes in
the ionization state (as we assume here), then the size is required to be 
a factor of $\sim$10$^3$
smaller, since density changes propagate at the sound speed.  Such a small
reflection region would be hard to reconcile with an edge-on torus 
geometry,
since the height of the inner edge of the torus is a factor of $\sim$10$^{3-4}$
times as large.  
As mentioned in section 4.2.2, changes in the ionization state can occur on
time scales of $\sim$2 hr, so variability due to photoionization is a
viable explanation.
}%
.
However, since the variation is only a factor of $\approx$1.6,
the size $l$ (treated as a 
diameter) could be a factor of several larger than the light travel time.
Hereafter, we write the size $l$ in terms of our fiducial value 0.1 pc
($l = l_{0.1}$ 0.1 pc) and note that $l_{0.1} \lapprox$ 2.

\subsubsection{Density}

\begin{figure}[hbn]
\epsscale{1.0}
\plotone{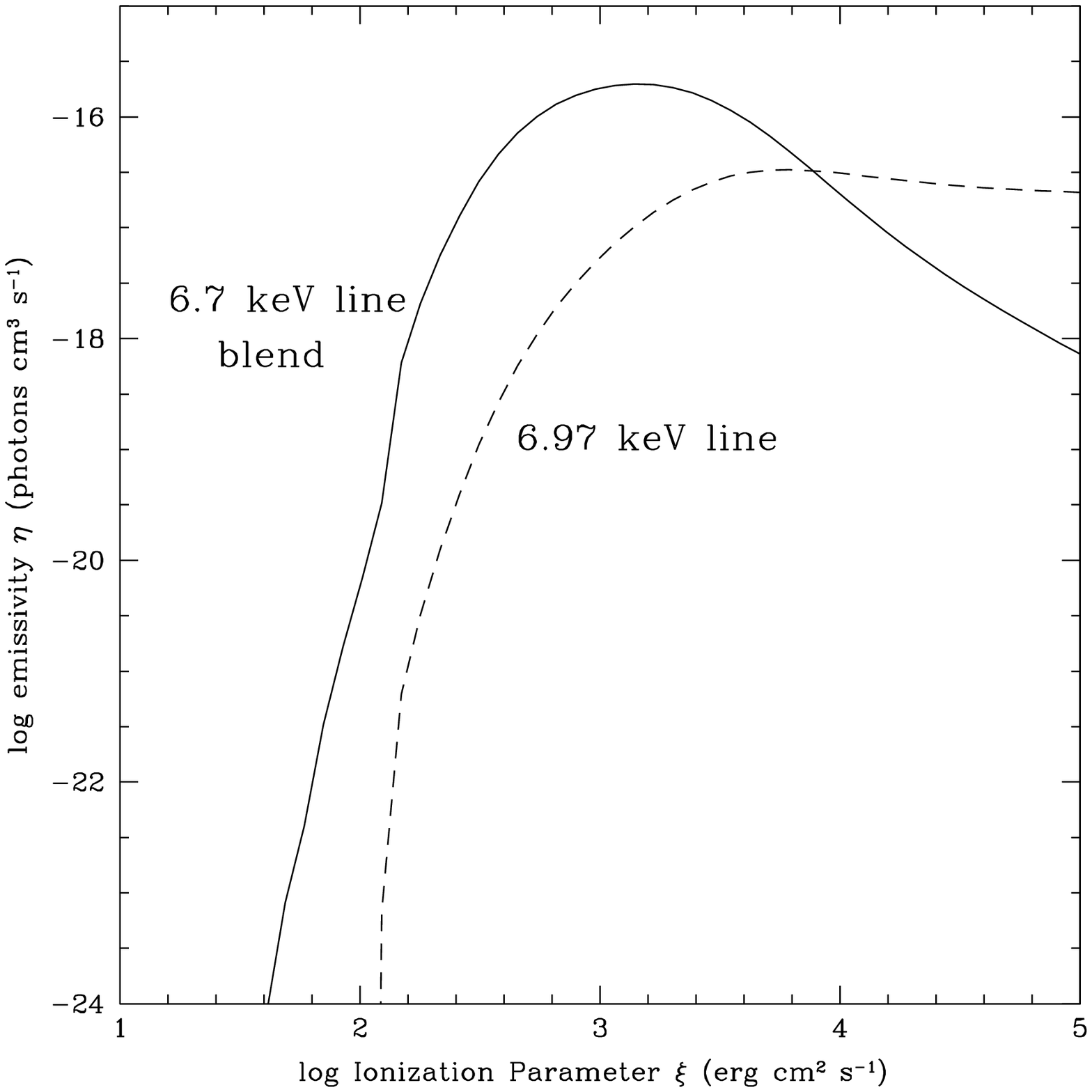}
\caption{
Emissivities $\eta$ of the 6.7 and 6.97 keV lines, as a function of the 
logarithm of the 
ionization parameter $\xi = L_0/nR^2$.
These values of $\eta$ were calculated
using XSTAR v1.45, assuming $T =$ 10$^7$~K.
\label{emiss}}
\end{figure}

A lower limit to the density can be estimated from the line flux $F$, since 
$$F = {{\eta \int n^2 dV} \over {4 \pi D^2}}   \eqno(1) ,$$
and the volume $V$ is limited by the size
$l$.  The combined emissivity $\eta$ for the three 6.7 keV lines 
(at 6.641, 6.669 and 6.683 keV) is plotted
in Figure~\ref{emiss}, 
along with the
emissivity for the 6.97 keV line.
For simplicity, we take 
$\eta = \eta_{-16}$ $\times$ (1.0 $\times$ 10$^{-16}$ 
photons~cm$^{3}$~s$^{-1}$),
where $\eta_{-16}$ is of order 1.
We next write
$\int n^2 dV = \bar{n}^2 \phi~l_{0.1}^{3}$ $\times$ (1.4 $\times$ 10$^{52}$ cm$^{-3}$), 
where the filling factor
$\phi$ (for a spherical volume) is less than 1.0, and the
cloud 
diameter $l_{0.1} \lapprox$ 2.
The 
line flux $F = f_{6.7}$ $\times$ (5.0 $\times$ 10$^{-5}$ photons~cm$^{-2}$~s$^{-1}$), where
$f_{6.7}$ is also of order unity ($f_{6.7} =$ 0.7 and 1.1$-$1.2 for model 2 for 
epochs 1 and 2, respectively).
For a distance $D = h_{75}^{-1}$ $\times$ (14.67 Mpc), we can then write
$$\bar n_{6.7}\approx 10^6 \left({f_{6.7}\over
\eta_{-16}\phi l_{0.1}^3}\right)^{1/2} h_{75}^{-1}\;{\rm cm^{-3}}$$
Since $\eta_{-16} <$ 1.8,
$\bar{n}_{6.7} \gapprox$ 10$^{5.5}$~cm$^{-3}$ to order of a few, 
since $\phi <$ 1 and $l_{0.1} <$ 2.  
Note that for R $\lapprox$ 0.1 pc, the ionization time scale for Fe~XXV,
$t_{ion} \sim 4 \pi R^2 E_{6.7~keV} / L_0 \sigma_{ion} \approx$ 2~hr,
for $L_0 =$ 10$^{43.5}$ erg~s$^{-1}$ and 
$\sigma_{ion}(Fe XXV) \approx$ 5 $\times$ 10$^{-20}$ cm$^2$ 
(e.g., Verner et al. 1996), so 
photoionization equilibrium is established much faster than the variability
time scale of 4 months.

\subsection{6.97 keV Line Emitting Region}

The 6.97 keV line is only strong for $\xi \gapprox$ 10$^{3.0}$ erg~cm~s$^{-1}$
(see Figure~\ref{emiss}), where
$\eta \approx$ 2.5 $\times$ 10$^{-17}$ photons~cm$^{3}$~s$^{-1}$.  
The density can also be estimated from the 6.97 keV line flux 
$F = ({\eta \int n^2 dV})/ ({4 \pi D^2}) \approx$ 1.0 $\times$ 10$^{-5}$
photons~cm$^{-2}$~s$^{-1}$, which requires 
$\bar n_{6.97} \gapprox 10^{5.9}/(\phi l_{0.1}^3)^{1/2} h_{75}\;{\rm cm^{-3}},$
very similar to the density
estimate of the 6.7 keV line emitting region.

\subsection{The 6.4 keV Line Emitting Region}

The line flux of the 6.4~keV line component is generally 10$-$20\% larger
in epoch~2.  However, the uncertainties in the line flux are larger than
this, so a constant value for both epochs is consistent with the data,
regardless of continuum model (see Table 3 and Figures 3 and 4).
The large 
equivalent width of the 6.4~keV component (EW $\sim$ 1~keV)
is almost certainly due to blockage of the direct 
X-ray continuum from the AGN, which would
exclude the possibility that
the line originates from the accretion disk.
Thus, the line emission likely originates from cold
gas further out (e.g., narrow-line clouds, or the inner edge of the torus;
Krolik, Madau, \& Zycki 1994, Ghisellini, Haardt, \& Matt 1994).

\subsection{Ionization Parameter and Size of the 6.7~keV and 6.97~keV Line Emitting Region}

\begin{figure}[hbn]
\epsscale{1.0}
\plotone{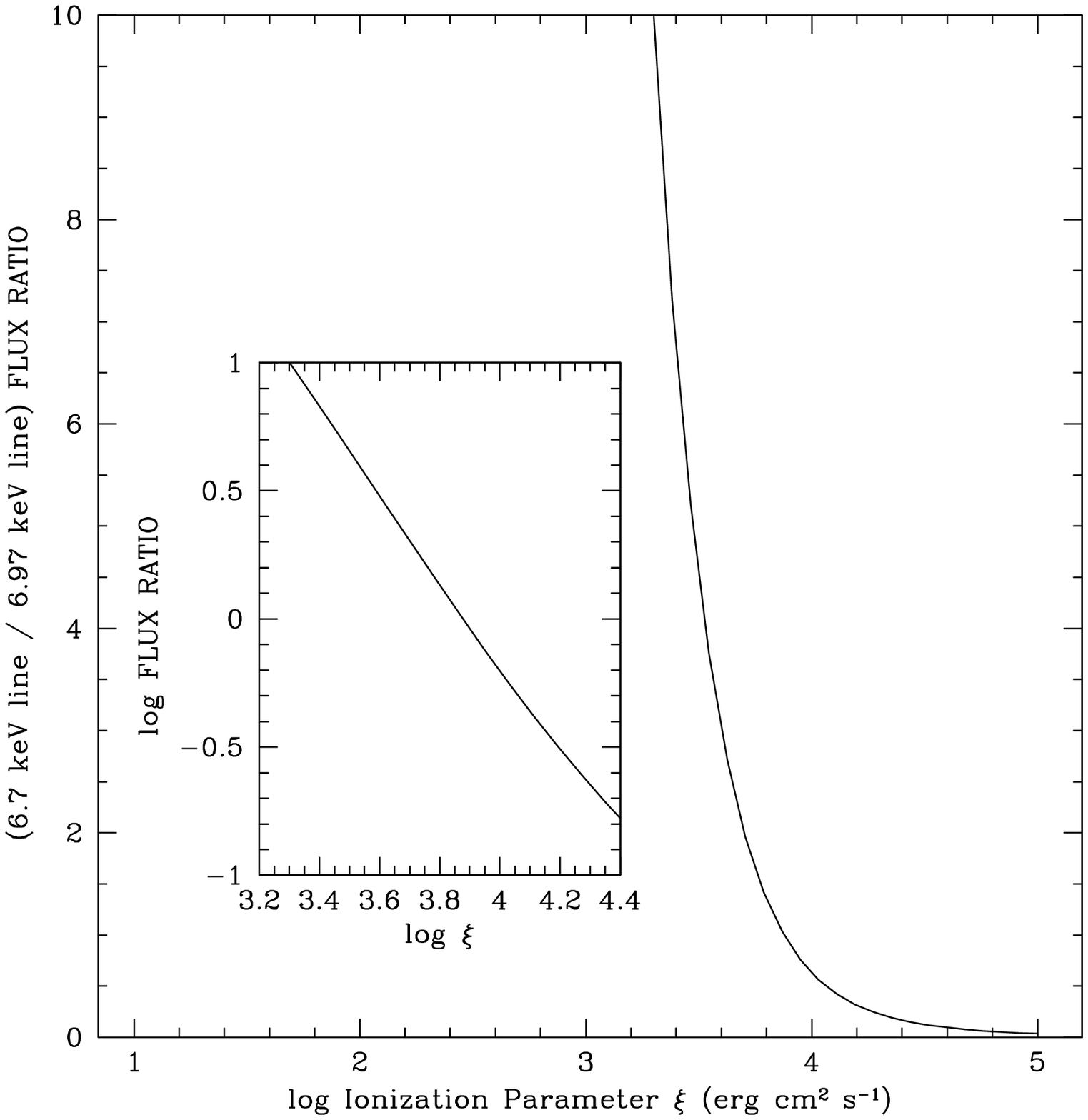}
\caption{
Expected line flux ratio F(6.7~keV)/F(6.97~keV), assuming both lines originate
from the same gas, as a function of the logarithm of the 
ionization parameter $\xi = L_0/nR^2$.
\label{emratio}}
\end{figure}

From the emissivity plot in Figure~\ref{emiss}, 
we can see that the 6.7~keV line emissivity is maximized for $\xi \sim 10^3$
erg~cm~s$^{-1}$, while the
emissivity of the 6.97~keV line rises with $\xi$ until it reaches
a flat maximum for log~$\xi \gapprox$ 3.5.
If only a single region
provides the line emission, then the observed line ratio can be used
to infer the value of $\xi$ in the emission region from the predicted
ratio of the two lines' emissivities.  Using Figure~\ref{emratio},
we find that the typical observed ratio $1.7 \leq F_{6.7}/F_{6.97} \leq 5.6$
(Table 3, variable power-law norm case) 
translates into 3.45 $\leq$ log~$\xi$ $\leq$ 3.75.

Because the emissivity of the 6.7~keV line {\it decreases} with
increasing $\xi$ in this range, whereas the emissivity of the 6.97~keV
line is roughly constant, we would predict on the basis of this
inference that decreases in continuum luminosity (for fixed density) would
lead to a rise in the 6.7~keV luminosity, but little change in the
6.97~keV line.  This is just what is observed.  Almost independent of
the different models used for line-fits (as shown in Table~3), $F_{6.7}$
{\it increased} by 60$-$70\%
from epoch 1 to epoch 2 while
$F_{6.97}$ is consistent with being constant.  The only model showing
a change in $F_{6.97}$ is the constant-flux, constant-continuum shape
model, which is a poor fit to the broad-band spectral data.

Quantitatively, the observed increase in $F_{6.7}$ between these two
epochs demands a decrease in $\xi$ (for this single-region model) of
$\approx$ 30\% (Figures~\ref{emiss} and \ref{emratio}).
Such a decrease is consistent with the change in
line ratio, the change in the absolute line fluxes, and the 
estimated 20\% drop in the continuum flux (section 3.3).

Given this confirmation of our estimate for $\xi$, we can estimate the
distance from the nucleus to the ionized-Fe emission line region 
from R $=$ (L/n$\xi$)$^{{{1}\over{2}}}$.  Using the
the emission measure estimate to write the density in
terms of the size of the emitting region, we have
$$ {{R}\over{l}} \approx 0.3 
  \left( { L_0 \over 10^{43.5}} \right)^{{{1}\over{2}}}
  \left( { \xi \over 10^{3.5}} \right)^{-{{1}\over{2}}}
  \phi^{{{1}\over{4}}}
  l_{0.1}^{-{{1}\over{4}}}.
\eqno(3)
$$
Since $R/l$ depends very weakly on $\phi$ and $l_{0.1}$, we
conclude that $R \sim l$.
In other words, the ionized
Fe line-emitting gas resides in a more-or-less volume-filling region
very close to the nucleus.
In summary, we estimate the following values for the 
physical parameters of the 6.7~keV and 6.97~keV
line emitting region: $l \lapprox$ 0.2~pc, $n \gapprox$ 10$^{5.5}$~cm$^{-3}$,
$log~\xi \approx$ 3.45$-$3.75, and $R \sim l$ (Table 4).

\subsection{The Continuum Reflection Zones}

As mentioned previously, the observed total 2$-$10 keV X-ray luminosity 
$L_{obs}$ is much smaller than the estimated intrinsic X-ray luminosity.
For example, when the data are fit with both a pure cold (neutral)
reflector component (XSPEC model {\sc pexrav})
and ionized reflection models
(models 2 or 4)\footnote{
The reduced $\chi^2$ values of these fits are $\sim$1.0, essentially the 
same as for 
the one-component ionized reflection models in Tables~\ref{tab2} 
and \ref{tab3}.
},   
$L_{obs}$(2$-$10~keV) $\approx$ 1.3 $\times$ 10$^{41}$ erg~s$^{-1}$
(model 2; 1.4 $\times$ 10$^{41}$ erg~s$^{-1}$ for model 4).
This corresponds to
a maximum X-ray reflection fraction $f_{refl} =$ 0.004 
(L$_0$/10$^{43.5}$)$^{-1}$.

As Pier et al. (1994) note, estimates of the reflected fraction for optical
light range from 0.001 to 0.05.
Miller, Goodrich \& Mathews (1991) estimate a value of 0.015 from measurements
of the broad H$\beta$ emission reflected from the NE knot.
Unless L$_0 \lapprox$ 10$^{43.0}$ erg~s$^{-1}$, it is
quite clear that 
the X-ray reflection fraction is, in general, smaller than the Miller et al. 
value.
In particular, from the joint (neutral$+$ionized reflection) fit, 
the neutral X-ray 
reflection component\footnote{
When computing 2$-$10 keV fluxes, we have included the 
6.21 and 6.4 keV 
line fluxes in the cold reflector flux, and the 6.7 and 6.97 keV line
fluxes in the warm reflector flux.
}
$L_N \approx$ 3.9 $\times$ 10$^{40}$ erg~s$^{-1}$
(model 2; 5.4 $\times$ 10$^{40}$ erg~s$^{-1}$ for model 4)
corresponds to a reflected fraction for the neutral reflector of 
$f_N =$ 0.001 (L$_0$/10$^{43.5}$)$^{-1}$,
and the ionized reflection component
$L_I \approx$ 10.0 $\times$ 10$^{40}$ erg~s$^{-1}$
(model 2; 7.3 $\times$ 10$^{40}$ erg~s$^{-1}$ for model 4),
corresponds to $f_I =$ 0.003 (L$_0$/10$^{43.5}$)$^{-1}$.

\subsubsection{The Ionized Reflector}

Since the 6.7/6.97~keV line emitting region is ionized to $\log\xi\sim 3.5$,
and this is close to the $\xi$ already estimated for the warm reflection
region (Table 2), we suppose that the ionized Fe lines originate from the
warm reflection region.  The luminosity reflected by that region is
$L_W = (\Omega/4\pi)_W (1 - e^{-\tau_W})L_0$, where the total
scattering depth is $\tau_W$.  Depending on details of the velocity and
ionization structure of the region, resonance line scattering could make
$\tau_W$ averaged over the 2 -- 10~keV band as much as a few times
greater than the Thomson depth $\tau_{Th} = n l \sigma_{Th} = 
0.2 l_{0.1}^{-1/2}$
(section 4.2; Krolik \& Kriss 1995; see also Kinkhabwala et al. 2002).  
We may then
turn around this last equation to estimate the allowed range for
$(\Omega/4\pi)_W$, where the limits correspond to $\tau_W =$ 0.14 
and $\tau_W = \infty$: 
$$ 
0.003 \left( { L_0 \over 10^{43.5}} \right)^{-1}
  \le \left( {\Omega \over 4\pi}\right)_W \le 
0.024 \left( { L_0 \over 10^{43.5}} \right)^{-1}.
\eqno(4)
$$
If the warm reflector is a single cloud and L$_0 =$ 10$^{43.5}$, this 
range in $(\Omega/4\pi)_W$ 
implies that the warm reflector has an opening half-angle of 
$\approx$6$-$18$^\circ$ with respect to the nucleus, and
$R_W/l_W \approx$ 1.5-4.5.  This range for $R/l$ is barely consistent with
our estimate that $R/l \sim$ 1 from section 4.5.  Furthermore, if
the cloud is spherical, the filling factor
$\phi_W =$ ${{1}\over{8}}(R_W/l_W)^{-3}$ is uncomfortably small 
($\lapprox$10$^{-2}$).
From equation $(3)$, this would require
$R/l$ to be less than unity, which is difficult geometrically and is 
inconsistent with the estimated range of
1.5-4.5.
A possible explanation is
that the AGN in NGC~1068 is intrinsically X-ray weak.  For example, if 
$L_0 \approx$ 10$^{43.0}$ erg~s$^{-1}$, $R_W/l_W \sim$ 1 is consistent with
the range of covering factors from equation $(4)$, so that $\phi_W \sim$ 1
is possible.
This scenario seems most consistent with the warm reflector properties.
We note that the scatter in the [O~III]/L$_X$ relationship is $\approx$0.25
(section 4.1), so we might have expected 
10$^{43.25} \lapprox$ L$_0$ $\lapprox$ 10$^{43.75}$ erg~s$^{-1}$.
Given the uncertainties in our estimate of L$_0$ (10$^{43.0}$),
we do not imply that NGC~1068 is 
{\it abnormally} weak for its [O~III] luminosity, but we do note that
10$^{43.0}$ erg$^{-1}$ is weaker than most Seyfert~1 nuclei
(e.g., Nandra et al. 1997).

\subsubsection{The Narrow-Line Region}

When the Narrow-Line Region (NLR) of NGC~1068 is imaged in polarized 
UV light with HST, the polarization is highest ($\approx$60\%) 
in the NE knot near ``cloud~B,'' which has a transverse size of 
$\approx$0.1\asec\  (7.1 pc) at a
projected distance of $\approx$3.6 pc
from the nucleus (Kishimoto 1999).  Although cloud~B is obviously a 
significant reflector, only $\sim$20\% of the [O~III] $\lambda$5007 line
flux (or [scattered] U-band emission) originates from cloud~B,
so that reflection from the rest of the NLR gas must be considered.

Using NIR/optical/NUV long-slit spectroscopic observations with STIS and the
HST, Kraemer \& Crenshaw (2000a) report three distinct line-emitting components
in the cloud~B aperture (0.1\asec\ $\times$ 0.2\asec).  
The two ``low-ionization'' ($\xi \lapprox$ 1)
components have column density N$_H \approx$ 1 $\times$10$^{21}$ cm$^{-2}$ and 
a combined covering factor $\Omega/4\pi \approx$ 0.0045.
Based on the photoionization modelling results of the whole NLR by
Kraemer \& Crenshaw (2000b), we estimate a total covering factor along
the STIS slit of $\sim$0.02.  By comparing the slit geometry with the
[O~III] $\lambda$5007 image (cf. Figure 1 of Crenshaw \& Kraemer 2000),
the covering factor for all of the low-ionization NLR gas could be
$\gapprox$10 times as large, i.e., appreciably near unity.  However, 
the Thomson optical depth for the low-ionization gas in the NLR is too small
to scatter much UV or X-ray light.
From Kraemer \& Crenshaw (2000a), N$_e \sim$ 1 $\times$ 10$^{20}$ cm$^{-2}$ 
within cloud~B.
The line-emitting gas in the HST [OIII] $\lambda$5007 FOS image of 
NGC~1068 is roughly contained in a 2.7\asec\ $\times$ 4.0\asec\ rectangular
region, so the linear dimension of the whole NLR
is $\sim$15$-$40 times that of the cloud~B aperture.  So, even if the net
electron column of the low-ionization gas in the NLR was a factor of 40 times 
that of 
cloud~B, $\tau_{Th} \lapprox$ 0.003.
Thus, only under very extreme (and probably unlikely) conditions would the
low-ionization gas in the NLR electron scatter an appreciable amount of
X-ray light, and would certainly not be able to reflect the observed fraction
of $\sim$1\% of scattered UV light.

The third gas component found by Kraemer \& Crenshaw (2000a) is a 
high-ionization ($\xi \sim$ 50) ``coronal'' component with  a column of
N$_H \approx$ 4 $\times$ 10$^{22}$ cm$^{-2}$ within the cloud~B aperture.
Kraemer \& Crenshaw (2000b) suggest that this coronal gas has a much
larger covering factor than the low-ionization gas.
If this coronal gas is as dense across the whole NLR,
the net electron column density 
could be $\gapprox$ a few 10$^{23}$ cm$^{-2}$, which is large enough to scatter 
a significant fraction of the incoming
UV light.
(We note, however, that Miller, Goodrich, \& Mathews 1991 favor Thomson
optical depths $<$ 0.1 for the optical scatterer.)

The ionization parameter of the coronal gas is low enough so that it acts as
a cold reflector to incoming X-rays from the AGN.
Based on Figure 2 of Capetti et al. (1997), 
the half-angle describing the NLR with respect to the nucleus is
approximately 45$^\circ$, so that $(\Omega/4\pi)_{NLR} =$ 0.15.  
If the coronal gas is optically thick, the expected
reflected X-ray emission 
is then L$_{NLR} = (\Omega/4\pi)_{NLR} a L_0,$
where $a$ is the X-ray albedo.
Reflection from neutral gas yields an X-ray albedo of 2.2\% in the 2$-$10 keV 
band (George \& Fabian 1991).  Based on XSPEC PEXRIV simulations,
we estimate a 25\% increase in the 2$-$10 keV albedo from $\xi=0$ to $\xi=50$,
or $a =$ 2.8\%.  This implies a maximum value for 
$L_{NLR} =$ 0.0042 L$_0 =$ 
1.3 $\times$ 10$^{41}$ erg~s$^{-1}$ (L$_0$/10$^{43.5}$).  
The total observed X-ray luminosity from neutral reflection 
L$_N$ is 3.9$-$5.4 $\times$ 10$^{40}$ erg~s$^{-1}$,
so it is possible that neutral reflection from coronal gas in the NLR could 
contribute significantly, although this would require L$_0$ to be 
$\lapprox$10$^{43}$ erg~s$^{-1}$, unless the 
coronal
gas is optically thin.

\subsubsection{The Torus}

\begin{figure}[hbn]
\epsscale{1.0}
\plotone{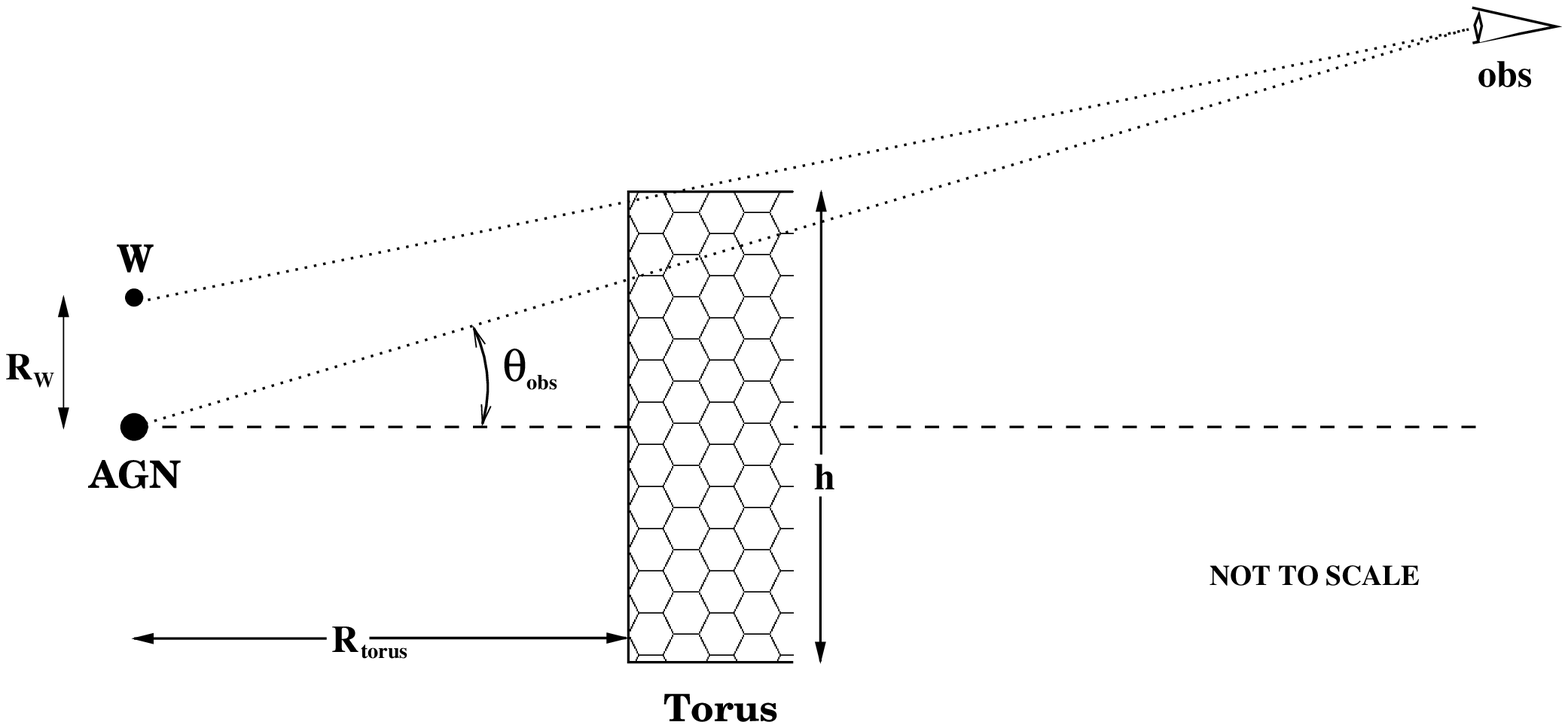}
\caption{
Cartoon of the proposed geometry of the warm reflector W and the dense,
obscuring torus (see section 4.6).  
The cross section of the torus is shown here as rectangular,
with full height $h$ and inner radius $R_{torus}$.  
The warm reflector $W$ is shown at radius $R_W$ from the AGN, and is 
positioned such that the line-of-sight to the observer just skims the
surface of the torus.  See Table 4 for properties of the warm reflector.
The inclination angle 
of the observer $\theta_{obs}$ is measured with respect to the equatorial 
plane of the torus. 
\label{warmrefltorus}
}
\end{figure}

    These data also allow us to draw a more complete picture of the obscuring
torus.  On the direct line of sight to the nucleus, we know
from the exceptionally weak 2--10~keV luminosity that the obscuration
must be at least Compton thick.  On the other hand, we can see the ionized
reflector only $\sim$0.1~pc away from the nucleus, so this nearby line
of sight must have a much smaller column density.  The column density
on this line of sight cannot, however, be zero because we see no evidence
for the warm reflector in optical light.  This column density, therefore,
may be $\sim 10^{21}$--$10^{22}$~cm$^{-2}$---enough so that its associated
dust will thoroughly block optical and UV photons, but not enough to
stop the X-rays we observe.   We therefore conclude that our line of
sight to the nucleus passes through a thicker part of the torus (near
the equatorial plane?) and that the column density drops sharply as
one moves away from that line of sight (or at least a hole).  We 
show a sketch of our
proposed geometry of the torus and
the warm reflector in Figure~\ref{warmrefltorus}.

    The covering fraction of the thickest part of the torus may also
be estimated.  In order to provide a significant ``cold reflection"
signal, the reflecting gas must be at least Compton thick.  Consequently,
it is only the thickest part of the torus that may contribute to this
part of the X-ray spectrum.  We have just seen that its transverse
size is $\sim 0.1$~pc.  Two pieces of evidence indicate that its
distance from the nucleus is $\simeq 0.5$~pc.  The VLBA images of
Gallimore et al. (1997) stretch to radial distances $\simeq 0.6$~pc.
Because the measured brightness temperature is $\sim 10^6$~K, the
source of radio flux cannot be part of the torus, making its inner
radius at least this large.  VLBA
maser observations, on the other hand, show that there is cool
molecular gas as close as $\simeq 0.5$~pc from the nucleus (Greenhill et al.).
On this basis, we suppose that $(\Omega/4\pi)_{torus} \simeq 0.1$.

    The torus's contribution to the ``cold reflected" luminosity is
the product of its covering fraction and its albedo.  Combining
the estimate of the previous paragraph with the cold reflection
albedo computed by George \& Fabian (1991) (2.2\% for the 2$-$10~keV range),
we would predict a torus contribution 
$\sim 7 \times 10^{40}$ (L$_0$/10$^{43.5}$), which is comparable to L$_N$,
our estimate of the X-ray reflection from neutral gas.
Therefore, cold reflection from the inner edge of the torus could also 
contribute significantly.

\subsection{Variability}

As a consistency check, we note that between epochs 1 and 2, the total
observed flux (and hence luminosity L$_{obs}$) 
decreased by a factor of 0.8, while L$_0$ and L$_W$ ($=$ L$_I$) are inferred
to have decreased by $\approx$0.2 dex, or by a factor of 0.63.  This can be 
accomplished if L$_I$/L$_{obs} =$ 0.54, assuming the cold reflector 
luminosity L$_N$ did not vary.  We note that for our joint
(neutral$+$ionized reflection) fits described in section 4.6, 
L$_I$/L$_{obs}$ is 0.72 for the optically thin warm reflector, and 0.58 for
the optically thick warm reflector.  Given the uncertainties of the
fluxes used to estimate these fractions, and the shortcomings of the XSPEC 
ionized reflection models, the fractions are consistent with each other.
The observational result from the joint fit implies that the
X-ray emission from the warm reflector contributes $\sim$60-70\% of the
total observed luminosity.

\subsection{A Consistent Model for the X-ray Reflectors}

Many potential problems discussed here can be alleviated if the intrinsic
X-ray luminosity of the AGN  L$_0 \lapprox$ 10$^{43.5}$ erg~s$^{-1}$.
If L$_0 \approx$ 10$^{43.0}$ erg~s$^{-1}$, the
covering factor and filling
factor of the warm reflector then become large enough to be consistent with 
equation $(3)$:
$R_W/l_W \sim$ 1.
This would also imply that the maximum X-ray reflection fraction 
f$_{refl} \lapprox$ 0.01, which is 
comparable to the reflection factor for optical and UV light.  
The best-fit spectral
models include a Compton reflection component, either from optically thick
ionized gas, or from optically thick neutral gas.  Therefore, if the optical
depth of the ionized gas in the 
warm reflector is low enough so that a Compton ``hump'' is not
present, reflection from optically thick neutral gas is needed.  The torus
can easily produce this cold reflection component, but the coronal gas in
the NLR may also contribute (if $\tau_{Th} \gapprox$ 1).

\section{Implications and Conclusions \label{implications}}

We propose that the 6.7/6.97 keV line emitting region
and the previously unidentified warm, ionized reflector
in NGC~1068 are indeed the same region.
Approximately 2/3 of the total observed 2$-$10 keV X-ray emission comes
from the warm reflector, which is located $\lapprox$0.2 pc from the AGN.
The geometrical and physical properties of the warm reflector
(Table~\ref{tabwarmrefl})
are quite different from those of the
optical/UV reflector (e.g. Miller et al. 1991).

The warm X-ray reflector should reflect
optical and UV light; however, no such emission is seen in HST images at the
location of the proposed nucleus (e.g., Capetti et al. 1995).  This suggests that
the dust optical depth toward the nucleus is high enough to block
the UV and optical light, but the equivalent X-ray column is not large enough
to block the harder X-ray emission.  Equivalent columns up to $\sim$10$^{22}$
cm$^{-2}$ are allowed when fitting the X-ray models to the ASCA/RXTE/BeppoSAX
data, so such a scenario is feasible from the X-ray point of view.  We can
achieve this scenario if the torus is viewed edge-on so that the 
line-of-sight toward the nucleus skims the edge of the torus
(see Figure~\ref{warmrefltorus}).

\begin{figure}[hbn]
\epsscale{1.0}
\plotone{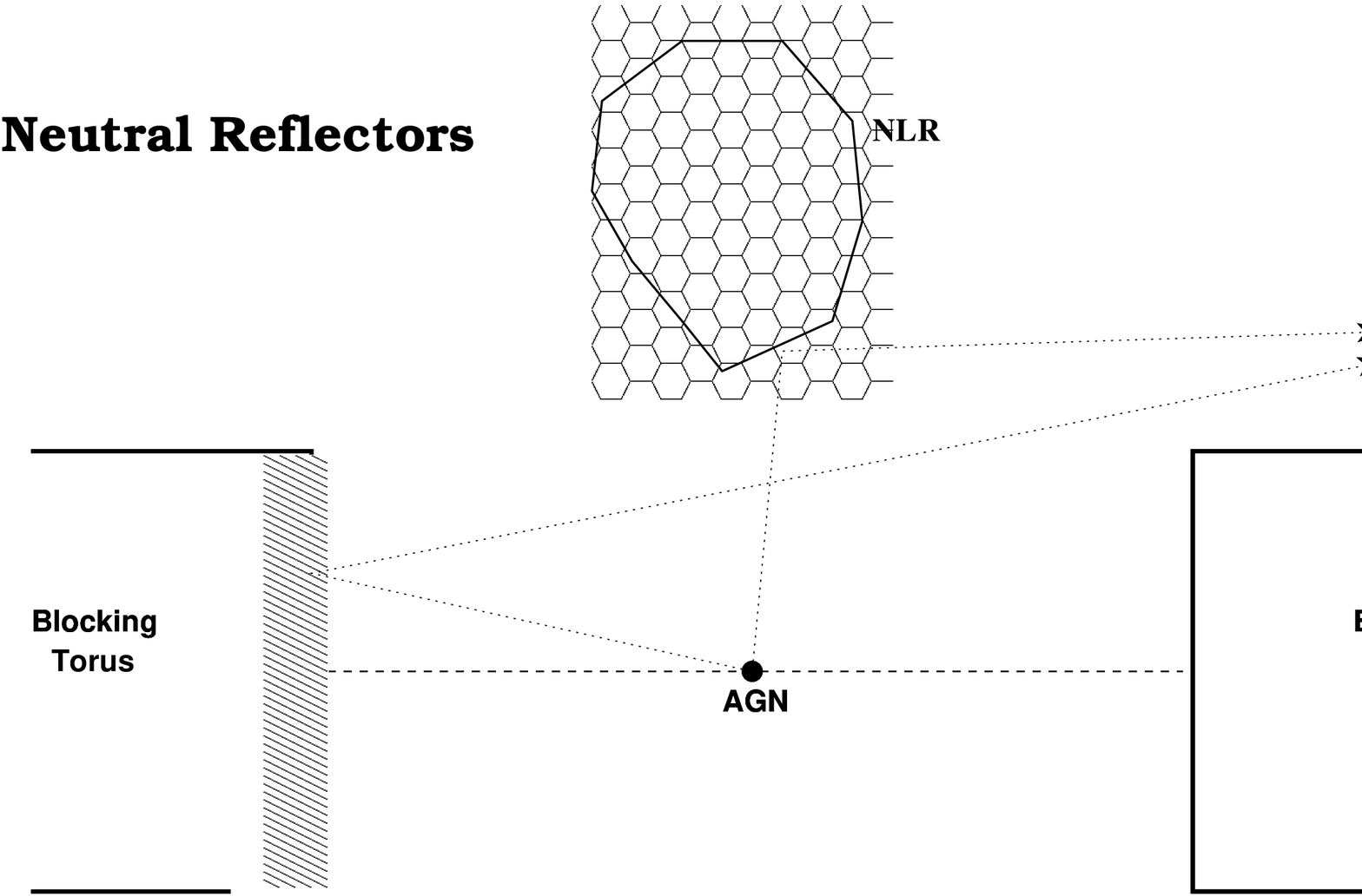}
\caption{
Cartoon of the two optically-thick `cold' reflection regions:
the inner surface of the dense, obscuring torus, and the optical reflection 
region (the inner NLR), located $\sim$30 pc from the AGN (Capetti et al. 1995).
The shaded region of the torus illustrates that only a fraction
($\lapprox$ 0.5) of the inner surface of the torus reflects X-rays 
toward the observer.  The coronal gas in the inner NLR has a moderately
low ionization parameter ($\xi \sim$ 50) and may be optically thick
to electron scattering, and therefore could also be
a ``cold'' X-ray reflector.
\label{neutralreflectors}
}
\end{figure}

We estimate that $\sim$ 1/3 of the total observed 2$-$10 keV X-ray emission 
is reflected by optically thick, neutral reflectors.
See Figure~\ref{neutralreflectors} for a cartoon illustrating the possible
reflection regions. 
If the coronal gas in the NLR is Compton thick, it is possible 
that all of the cold reflection component is from the NLR.
We estimate a covering factor
of $\approx$0.15 for the inner NLR, and $\sim$0.1 
for the inner surface of the torus.  Thus, either of these two regions could
contribute significantly to the observed cold reflection component.
Likewise, 
the 6.4 keV Fe emission line could originate from either of these regions.

VLBA observations of NGC~1068 (Gallimore et al. 1997) show 
that the size, radius and density of the radio-emitting clumps at the 
inner edge of the torus 
($R \approx$ 0.6~pc, height $h \lapprox$ 0.1 pc) 
are very similar to those of our proposed warm reflector.
If the warm
reflector does indeed have a density $\gapprox$10$^6$~cm$^{-3}$ and a
spherical volume of diameter $\sim$0.1~pc,
and is at temperature of $\sim$0.1-1 keV, 
the emission measure will be large enough to see thermal bremsstrahlung 
radio emission (cf. Gallimore et al. 1997).
However, since the emission 
$\propto$~n$^2$, values of $n$ below a few $\times$ 10$^5$ 
(e.g., if D(NGC~1068) $<$ 14.67 Mpc)
would force the
bremsstrahlung luminosity below
Gallimore et al.'s detection limit.
It is worth noting that since
the morphology of the radio clumps of S1 are not co-linear (as one would
expect from an unwarped,
edge-on disk/torus), it is quite possible that one of the
clumps is 
in the Gallimore et al. radio image is the primary warm X-ray reflector.

How does our proposed warm reflector region fit in with the AGN paradigm?
Krolik \& Kriss (1995) suggest that the ``warm absorbers'' seen in type~1
AGN may be associated with the warm reflection region.  Typical
ionization parameters associated with the warm absorbers are $\xi \sim$ 
10$-$100 erg~cm~s$^{-1}$ (e.g., Reynolds 1997, Kaspi et al. 2001), 
which is much lower than the values we find ($\xi \sim$ 10$^3$).  
In principle, there could be clouds with different densities and 
mean ionization parameters in each of these regions, which could be
responsible for the warm absorbers see in AGN X-ray spectra.
One possibility is that the more 
highly ionized, inner regions of the warm reflector clouds
reflect X-rays toward our line of sight, while the more weakly ionized, 
outer regions of the clouds cause the warm, ionized edges.

We suggest the existence of ionized, warm reflector gas clouds distributed
within a radius of $\sim$0.2 pc.
The highly ionized gas in this region produces an
Fe~K edge near 8.6 keV in the reflected nuclear spectrum, as well as both
the observed 6.7~keV and 6.97~keV emission lines.
Cold reflection, required to model the observed Compton reflection hump,
originates from optically thick ``cold'' gas in the inner NLR
and/or from the inner surface of the torus.
Our data and proposed models are most consistent with an intrinsically X-ray
weak AGN with L$_0 \approx$ 10$^{43.0}$ erg~s$^{-1}$.

\acknowledgments

We thank Tahir Yaqoob, Jack Gallimore, Steve Kraemer, 
Mike Crenshaw and the anonymous
referees for helpful discussions, and Tim Kallman for help with XSTAR.  
We also thank Andy Ptak for use of his TCL scripts for computing 
uncertainty ranges of XSPEC model parameters.

\clearpage

\clearpage

\begin{deluxetable}{lllccl}
\tablecaption{Log of X-ray Spectral Observations of NGC~1068 \label{tab1}}
\tablewidth{0pt}
\tablehead{
\colhead{Epoch} & \colhead{Dataset} &
\colhead{ID\tablenotemark{a}} & \colhead{Observation Dates} &
\colhead{Exp. Time} & \colhead{Detector\tablenotemark{b}} \\
\colhead{} & \colhead{} & \colhead{} & \colhead{(YYMMDD)} &
\colhead{(ks)} & \colhead{} \\
}
\startdata
0 & ASCA1 & 70011000 & 930724$-$930725 & 21.8 & S0 \\
1 & ASCA2 & 74034000 & 960815$-$960818 & 93.8 & S0 \\
1 & RXTE  & 10322    & 960816$-$960819 & 48.3 & PCA \\
2 & SAX1  & 50047001 & 961230$-$970103 & 100.8 & MECS \\
3 & SAX2  & 50047002 & 980111$-$980112 & 37.3 & MECS \\
 \enddata

\tablenotetext{a}{Data ID: sequence number or proposal number}
\tablenotetext{b}{Detector for which exposure time is quoted}

\end{deluxetable}
\clearpage

\begin{deluxetable}{llcccl}
\footnotesize
\tablecaption{Epoch 1 Simple Continuum Model Fits: ASCA2$+$RXTE$+$SAX1/PDS \label{tab2}}
\tablewidth{0pt}
\tablehead{
\multicolumn{2}{l}{Model} & 
\colhead{$\Gamma$} & 
\colhead{R} & \colhead{$\chi^2$/dof} & \colhead{Notes} \\
}
\startdata
\multicolumn{6}{l}{Fe~K Complex fit Empirically with Four Fixed Gaussian Lines\tablenotemark{a}} \\
1 & Abs.P.L.  & 1.68 & {...} & 375.6/322 & N$_H =$ 1.0 $\times$ 10$^{23}$~cm$^{-3}$ \\
2 & P.L.$+$Edge & 1.04 & {...} & 343.8/321 & E$_{edg} =$ 8.60 keV, $\tau_{edg} =$ 0.64 \\
3 & N.Refl.     & 1.44 & 1.19  & 401.2/322 & \\ 
4 & I.Refl.      & 1.56 & 4.53  & 312.6/321 & $\xi =$ 1798 \\
\tableline
\multicolumn{6}{l}{Fe~K Complex fit with Four Narrow Gaussian Lines\tablenotemark{b}} \\
1 & Abs.P.L. & 1.43 & {...} & 370.2/318 & N$_H =$ 4.97$_{-3.28}^{+3.11}$ $\times$ 10$^{22}$~cm$^{-2}$ \\ 
  &     & 1.29-1.57 &       &           & N$ =$ 7.96$_{-2.41}^{+3.29}$ $\times$ 10$^{-4}$ photons~keV$^{-1}$~cm$^{-2}$~s$^{-1}$ \\ 
2 & P.L.$+$Edge & 1.06 & {...} & 333.3/317 & E$_{edg} =$ 8.63$_{-0.21}^{+0.23}$ keV, $\tau_{edg} =$ 0.54$\pm$0.14 \\
  &     & 1.01-1.13    &       &           & N$ =$ 3.67$_{-0.35}^{+0.45}$ $\times$ 10$^{-4}$ photons~keV$^{-1}$~cm$^{-2}$~s$^{-1}$ \\ 
3 & N.Refl.     & 1.46 & 1.48  & 366.9/318 & \\ 
  &      & 1.31-1.66 & 0.52-3.53 &         & N$ =$ 6.31$_{-1.15}^{+1.70}$ $\times$ 10$^{-4}$ photons~keV$^{-1}$~cm$^{-2}$~s$^{-1}$ \\ 
4 & I.Refl.     & 1.54 & 4.07  & 313.4/317 & $\xi =$ 1840$_{-1020}^{+3160\tablenotemark{c}}$ erg~cm~s$^{-1}$ \\
  &      & 1.36-1.69 & 2.69-7.11 &         & N$ =$ 4.75$_{-1.03}^{+0.96}$ $\times$ 10$^{-4}$ photons~keV$^{-1}$~cm$^{-2}$~s$^{-1}$ \\ 
\enddata

\tablenotetext{a}{
For the ``empirical'' four-Gaussian fits, we have modelled the Fe~K line 
complex by fixing the parameters of the Gaussian components to the following
values (energy, line width $\sigma$, line flux):
(6.21 keV, 10.0 eV, 1.36 $\times$ 10$^{-5}$ photons~cm$^{-2}$~s$^{-1}$); 
(6.38 keV, 47.9 eV, 4.75 $\times$ 10$^{-5}$);
(6.59 keV, 7.3 eV, 2.17 $\times$ 10$^{-5}$), and 
(6.84 keV, 90.0 eV, 2.62 $\times$ 10$^{-5}$).
}
\tablenotetext{b}{
For these fits, we have fixed the central energies of the four Gaussian lines
to 6.21, 6.40, 6.70, and 6.97 keV.  Line line widths were fixed at 
$\sigma =$ 0, except for the 6.21 keV line, for which we fixed $\sigma$ at 
150 keV.
The line fluxes for the four lines are listed under ``Epoch 1'' in Table 3.
}
\tablenotetext{c}{
The upper limit to $\xi$ quoted here is the maximum value allowed by the PEXRIV
model: 5000 erg~cm~s$^{-1}$. 
}
\end{deluxetable}
\clearpage

\begin{deluxetable}{lccccccccc}
\footnotesize
\tablecaption{Fe~K line fluxes for Epochs 1 and 2 (Narrow-Line Model)
  \label{tab3}}
\tablehead{
\multicolumn{5}{l}{Model and Parameters\tablenotemark{a}} &
\multicolumn{4}{c}{Line Fluxes (10$^{-5}$ photons~cm$^{-2}$~s$^{-1}$)} &
\colhead{$\chi^2$/dof} \\ 
\colhead{} & \colhead{} & \colhead{} & \colhead{} & \colhead{} &
\colhead{6.21} & \colhead{6.4} & \colhead{6.7} & \colhead{6.97} \\
\colhead{} & \colhead{} & \colhead{} & \colhead{} & \colhead{} &
\colhead{(keV)} & \colhead{(keV)} & \colhead{(keV)} & \colhead{(keV)} \\
}
\startdata
& \multicolumn{9}{l}{Epoch 1\tablenotemark{b}~~~ (see Table 2) } \\
1  &  & && & 2.38 & 4.76 & 3.39 & 1.79 &  \\
         &  && & & 1.79-2.98 & 4.10-5.41 & 2.86-3.94 & 1.38-2.21 \\
2  & && & & 2.38 & 4.67 & 3.40 & 1.52 & \\
            & &&  &  & 1.82-2.93 & 4.03-5.32 & 2.86-3.93 & 1.11-1.94 \\
3  &  & &&       & 2.66 & 4.64 & 3.38 & 1.98 & \\
         &    &&       &     & 2.10-3.21 & 3.99-5.29 & 2.84-3.92 & 1.57-2.39 \\
4  &  & &&       & 2.02 & 4.71 & 3.34 & 1.03 & \\
         &    &&       &     & 1.47-2.59 & 4.07-5.36 & 2.82-3.89 & 0.61-1.46 \\
\tableline
& \multicolumn{9}{l}{Epoch 2\tablenotemark{c}~~~ 
  (Epoch 2 Continuum Model Fixed at Epoch 1 Parameters -- } \\
& \multicolumn{9}{l}{~~~~~~~~~~~~~~~~~~~~CONSTANT power-law norm)} \\
1 &  &&&   & 0 & 4.45 & 5.84 & 0.18 & 555.1/420 \\
         &   &&   & & $<$0.23 & 3.68-5.22 & 4.77-6.75 & $<$0.90 \\
2  &&&  &  & 0 & 4.45 & 5.96 & 0 & 501.4/420 \\
         &   &&   &      & $<$0.24 & 3.73-5.21 & 4.90-6.69 & $<$0.70 \\
3 &  &&&  & 0 & 4.78 & 5.63 & 0.52 & 529.9/420 \\
         &   &&   &      & $<$0.27 & 4.00-5.55 & 4.56-6.70 & $<$1.23 \\
4  &  &&&  & 0 & 4.25 & 5.57 & 0 & 502.9/420 \\
         &   &&   &      & $<$0.19 & 3.52-4.96 & 4.82-6.30 & $<$0.31 \\
\tableline
& \multicolumn{9}{l}{Epoch 2\tablenotemark{c}~~~ 
  (Epoch 2 Continuum Model Fixed at Epoch 1 Parameters -- } \\
& \multicolumn{9}{l}{~~~~~~~~~~~~~~~~~~~~VARIABLE power-law norm)} \\
1 & N=6.45     &&&&  0       & 5.51       & 5.28      & 1.13      & 480.5/418 \\
  & 5.89-6.74  &&&&  $<$0.52 & 4.62-6.32  & 4.20-6.36 & 0.39-1.88 \\
2 & N=2.98     &&&&  0       & 5.50       & 5.39      & 0.96      & 429.5/417 \\
  & 2.56-3.11  &&&&  $<$0.57 &  4.23-6.30 & 4.01-6.47 & 0.22-1.70 \\
3 & N=5.17     &&&&  0       & 5.73       & 5.13      & 1.36      & 465.2/418 \\
  & 4.38-5.40  &&&&  $<$0.64 &  4.23-6.53 & 3.62-6.21 & 0.57-2.10 \\
4 & N=3.82     &&&&  0       & 5.14       & 5.65      & 0.45      & 420.2/419 \\
  & 3.41-3.99  &&&&  $<$0.47 &  4.28-5.95 &  4.56-6.71& $<$1.20   \\
\tableline
\\
\\
\\
\\
\\
\\
& \multicolumn{9}{l}{Epoch 2\tablenotemark{c}~~~ 
  (Epoch 2 Continuum Model Parameters Allowed to Vary)} \\
\\
1  & $\Gamma$=1.55 & & & & 0 & 5.55 & 5.24 & 1.22 & 482.0/419 \\
         & 1.53-1.59     & & & & $<$0.50 & 4.65-6.34 & 4.17-6.33 & 0.46-1.95 \\
1 & N$_H$=10.0    & & & & 0 & 4.99 & 5.52 & 0.64 & 503.8/419 \\
         & 8.7-11.3      & & & & $<$0.34 & 4.20-5.77 & 4.43-6.59 & $<$1.37 \\
1  & $\Gamma$=1.56 & N$_H$=4.56 & & & 0 & 5.57 & 5.23 & 1.24 & 482.0/418 \\
         & 1.51-1.61    & 2.34-6.84  &  & & $<$0.51 & 4.67-6.37 & 4.16-6.32 & 0.48-1.99 \\
1  & $\Gamma$=1.24 & N$_H$=0 & N=3.89  & & 0 & 5.72 & 5.24 & 1.25 & 472.5/417 \\
         & 1.12-1.37    & $<$1.99    & 3.11-5.13 & & $<$0.67 & 4.70-6.52 & 4.16-6.32 & 0.50-1.99 \\
\tableline
2\tablenotemark{d} & $\Gamma$=1.18 & & & & 0 & 5.51 & 5.37 & 1.02 & 430.3/419 \\
         & 1.15-1.22        & & & & $<$0.54 & 4.59-6.31 & 4.29-6.45 & 0.28-1.77 \\
2\tablenotemark{d} & $\tau_{edg}$=0.84 & & & &  0 & 4.45 & 5.96 & 0 & 496.6/419 \\
         & 0.60-1.12             & & & & $<$0.24 & 3.73-5.22 & 4.90-6.69 & $<$0.70 \\
2\tablenotemark{d} & $\Gamma$=1.14 & $\tau_{edg}$=0.41 & N=3.40 & & 0 & 5.55 & 5.35 & 1.04 & 428.1/416 \\
         & 1.03-1.26 & 0.16-0.70 & 2.53-4.16                & & $<$0.58 & 4.51-6.35 & 4.24-6.44 & 0.29-1.79 \\
\tableline
3  & $\Gamma$=1.58 & & & & 0 & 5.72 & 5.00 & 1.46 & 462.5/417 \\
         & 1.55-1.67     & & & & $<$0.65 & 4.36-6.60 & 3.59-6.16 & 0.72-2.21 \\
3  & $\Gamma$=1.68 & R=2.7 & & & 0 & 5.15 & 4.51 & 1.33 & 462.8/417 \\
         & 1.50-1.88     & $<$10.1 & & & $<$0.68 & 4.25-6.65 & 3.51-6.15 & 0.64-2.24 \\
3\tablenotemark{e}  & $\Gamma$=1.68 & R=5.0 & N=6.30 & & 0 & 5.58 & 5.07 & 1.33 & 465.5/417 \\
\tableline
         & 1.40-1.95     & 1.4-12.9 & 4.47-8.03 & & $<$0.53 & 4.51-6.40 & 3.93-6.15 & 0.59-2.07 \\
4 & $\Gamma$=1.65 & & & & 0 & 5.16 & 5.61 & 0.53 & 422.6/419 \\
         & 1.63-1.69     & & & & $<$0.45 & 4.31-5.96 & 4.53-6.69 & $<$1.28 \\
4 & $\Gamma$=1.49 & R=2.41 & $\xi$=300 & & 0 & 5.42 & 5.45 & 0.82 & 414.9/417 \\
         & 1.38-1.63     & 0.94-5.41 & $<$2130 &  & $<$0.56 & 4.44-6.27 & 4.00-6.55 & $<$1.97 \\
4\tablenotemark{e} & $\Gamma$=1.62 & R=3.92 & $\xi$=0   & N=5.95 & 0 & 5.58 & 5.12 & 1.26 & 414.0/416 \\
         & 1.38-1.82     & 1.28-7.97 & $<$4550 & 3.70-7.55 & $<$0.41 & 5.00-6.20 & 4.34-6.31 & 0.20-2.10 \\
\tableline
\\
\\
\\
\\
\\
\\
\\
\enddata

\tablenotetext{a}{ 
Model number (see Table~2), and parameters that were allowed to vary during the joint fit:
$\Gamma =$ photon index of the power-law, N$_H$[10$^{22}$~cm$^{-2}$]$~=$ effective Hydrogen
absorption column, R $=$ reflection scaling factor R (fraction of reflected
emission that reaches the observer divided by 
the seed emission that reaches the observer), 
$\tau_{edg} =$ optical depth of $\sim$8.5 keV absorption edge feature,
N[10$^{-4}$~photons~keV$^{-1}$~cm$^{-2}$~s$^{-1}$]$~=$ XSPEC normalization constant (at 1 keV) for power-law model,
and $\xi$[erg~cm~s$^{-1}$]$~=$ PEXRIV ionization parameter.
} 
\tablenotetext{b}{
Line fluxes for epoch~1 data, from fits to models in Table~2.  
See Table~2 for values of other model parameters.
The equivalent widths for the absorbed power-law model
(GIS2 detector) are 
410$\pm$100, 860$\pm$120, 650$^{+110}_{-100}$, and 370$\pm$85 eV,
for the 6.21, 6.4, 6.7, and 6.97 keV line components.
}
\tablenotetext{c}{Line fluxes for epoch~2 data.  These fluxes were obtained
by fitting the epoch~1 data together with the epoch~2 data.  
During the fit, all model parameters for the epoch~1 data were fixed at the 
values listed under ``Epoch~1'' in the table.
The continuum model parameters for the 
epoch~2 data (SAX1 MECS and PDS) were either fixed to those of the epoch~1 
models, or allowed to vary in the fit. 
Equivalent widths for the absorbed power-law model
(MECS detector) are
0$^{+50}$, 800$\pm$140, 1120$^{+170}_{-210}$, and 37$^{+148}_{-37}$ eV,
for the 6.21, 6.4, 6.7, and 6.97 keV line components.
}
\tablenotetext{d}{
  E$_{edg}(MECS)$ fixed at Epoch 1 value to get unique fit.
}
\tablenotetext{e}{
XSPEC v.11 models PEXRIV and PEXRAV use different tables for computing 
opacities (A. Zdziarski 2002, priv. comm.), so PEXRIV with $\xi =$ 0 is
unfortunately numerically different from PEXRAV.  Therefore, the $\chi^2$
values for these two fits are not equal.
}
\end{deluxetable}

\begin{deluxetable}{llcl}
\tablecaption{Ionized Warm Reflector Properties \label{tabwarmrefl}}
\tablewidth{0pt}
\tablehead{
\multicolumn{2}{c}{Property} & \colhead{Value} & \colhead{Unit} \\
}
\startdata
$l_W$ & Diameter & $\lapprox$ 0.2 & pc \\
$n_W$ & Density  & $\gapprox$ 10$^{5.5}$ & cm$^{-3}$ \\
$log~\xi_W$ & Ionization Parameter & 3.45$-$3.75 & $log$(erg~cm~s$^{-1}$) \\
$R_W$  & Radius (from AGN) & $\sim l_W$ &  \\
 \enddata

\end{deluxetable}
\clearpage

\end{document}